\documentclass{jfm}

\usepackage{graphicx}
\usepackage{newtxtext}
\usepackage{newtxmath}
\usepackage{bm}
\usepackage{natbib}
\usepackage{hyperref}
\hypersetup{
    colorlinks = true,
    urlcolor   = blue,
    citecolor  = black,
}

\usepackage{makecell}
\usepackage{subcaption}

\usepackage{array}
\newcommand{\PreserveBackslash}[1]{\let\temp=\\#1\let\\=\temp}
\newcolumntype{C}[1]{>{\PreserveBackslash\centering}p{#1}}
\newcolumntype{R}[1]{>{\PreserveBackslash\raggedleft}p{#1}}
\newcolumntype{L}[1]{>{\PreserveBackslash\raggedright}p{#1}}

\newcommand{\RomanNumeralCaps}[1]
\newcommand{\bi}{\mathsfbi}

\title{Autophoretic skating along permeable surfaces}

\author{G\"{u}nther Turk\aff{1}
  \corresp{\email{guenther.turk@princeton.edu}},
  Rajesh Singh\aff{2}
 \and Howard A. Stone\aff{3}}

\affiliation{\aff{1}Princeton Materials Institute, Princeton University, Princeton, NJ, 08544, USA
\aff{2}Department of Physics, IIT Madras, Chennai 600036, India
\aff{3}Department of Mechanical and Aerospace Engineering, Princeton University, Princeton, NJ, 08544, USA}

\begin{document}
\maketitle

\begin{abstract}
    The dynamics of self-propelled colloidal particles are strongly influenced by their environment through hydrodynamic and, in many cases, chemical interactions.
    We develop a theoretical framework to describe the motion of confined active particles by combining the Lorentz reciprocal theorem with a Galerkin discretisation of surface fields, yielding an equation of motion that efficiently captures self-propulsion without requiring an explicit solution for the bulk fluid flow. 
    Applying this framework, we identify and characterise the long-time behaviours of a Janus particle near rigid, permeable, and fluid-fluid interfaces, revealing distinct motility regimes, including surface-bound skating, stable hovering, and chemo-hydrodynamic reflection. 
    Our results demonstrate how the solute permeability and the viscosity contrast of the surface influence a particle's dynamics, providing valuable insights into experimentally relevant guidance mechanisms for autophoretic particles. 
    The computational efficiency of our method makes it particularly well-suited for systematic parameter sweeps, offering a powerful tool for mapping the phase space of confined active particles and informing high-fidelity numerical simulations.
\end{abstract}

\section{Introduction}
\label{sec:introduction}
The interaction of fluid-borne particles with boundaries is a fundamental problem in low-Reynolds-number hydrodynamics, with relevance to biological locomotion \citep{brennen1977}, colloidal self-assembly \citep{palacci2013living}, and phoretic transport \citep{kreuterTransportPhenomenaDynamics2013}. 
In particular, colloidal particles that are self-propelled by interfacial forces, such as Janus particles, exhibit complex dynamics near boundaries, governed by a balance of hydrodynamic and phoretic interactions \citep{andersonColloidTransportInterfacial1989,uspalSelfpropulsionCatalyticallyActive2015,mozaffariSelfdiffusiophoreticColloidalPropulsion2016}.

A wide range of theoretical and numerical approaches have been employed to describe the dynamics of confined active particles, each with distinct advantages depending on the system geometry and level of approximation. Boundary element methods (BEM) offer accurate results for complex system geometries, but come at a high computational cost \citep{youngren1975stokes,power1987second,pozrikidisBoundaryIntegralSingularity1992}. Multipole methods are computationally efficient and provide analytical insights into particle motion, but are exact only for point singularities, leading to convergence issues for finite-sized particles \citep{blakeFundamentalSingularitiesViscous1974}. For spherical particles in simple geometries, methods based on bispherical coordinates largely avoid such truncation errors and yield semi-analytical solutions to high accuracy. However, their extension to more complex systems remains challenging  \citep{BRENNER1961242,leeMotionSpherePresence1980,papavassiliouExactSolutionsHydrodynamic2017}.
In this work, we take an alternative approach by leveraging the Lorentz reciprocal theorem, combined with a Galerkin discretisation of the emerging surface fields, to derive a governing equation for the motion of a confined active particle. Being based on the boundary integral representation of the Stokes equation, our method does not require a detailed knowledge of the flow field around the particle, thereby obviating the need for solving the underlying equations in the bulk. This also has the advantage that our method is readily applicable to any system for which a Green's function for the Stokes flow is known.

Building on this framework, we study the long-time dynamics of a catalytically active Janus particle near a variety of surfaces, including chemically permeable solids and fluid-fluid interfaces. We construct phase diagrams that reveal the emerging stable dynamical states typically predicted and observed for Janus particles near planar surfaces \citep{dasBoundariesCanSteer2015,dasFloorCeilingSlidingChemically2020,uspalSelfpropulsionCatalyticallyActive2015,simmchenTopographicalPathwaysGuide2016}, such as surface-bound skating, hovering, or reflection by the boundary. Our results show how chemical and hydrodynamic interactions determine the particle’s equilibrium tilt angle and motion, with the permeability and the viscosity contrast of the bounding surface significantly altering the observed behaviours \citep{palaciosGuidanceActiveParticles2019}. A key advantage of our method is its ability to efficiently explore a broad range of particle properties and environmental conditions, enabling the systematic identification of distinct motility regimes that would be challenging to capture with fully numerical methods.

The rest of the paper is organised as follows. In section \ref{sec:theory} we derive the dynamics of a confined active particle by combining two approaches to low-Reynolds-number hydrodynamics: the Lorentz reciprocal theorem and a Galerkin discretisation of the boundary integral representation of Stokes flow. In section \ref{sec:autophoresis} we then apply this method to identify and categorise the dynamics of a catalytically active Janus particle near a variety of surfaces and interfaces characterised by their solute permeabilities and viscosity contrasts. We conclude  in section \ref{sec:discussion} with a brief discussion of the results, contrasting them to previous works on steady states of confined Janus particles, and potential future applications thereof.

\section{Reciprocal relation for a confined active particle}
\label{sec:theory}
In this section we derive the governing equations for the dynamics of a confined active particle. 
We first use the Lorentz reciprocal theorem to establish a direct connection between the particle's activity and its rigid body motion, in principle taking into account all hydrodynamic interactions with its environment.
Using a Galerkin discretisation of the involved surface fields, we then provide a systematic way of making this connection explicit, expressing the result in terms of known quantities, so-called propulsion tensors. 

Denoting the centre of mass of the particle by $\bm{x}_0$ and defining $\bm{r}=\bm{x}-\bm{x}_0$, the fluid velocity distribution on the surface of an active particle $S$ is modelled by \citep{lighthillSquirmingMotionNearly1952,blakeSphericalEnvelopeApproach1971}
\begin{equation}
    \bm{v}(\bm{x})=\bm{V}+\bm{\Omega}\times\bm{r}+\bm{v}_s(\bm{x})
    \quad\text{for}\quad \bm{x}\in S.
    \label{eq:slip}
\end{equation}
Here, $\bm{V}$ and $\bm{\Omega}$ are the linear and angular velocities of the particle, respectively. 
The slip velocity $\bm{v}_s$ arises from interfacial forces, such as gradients in chemical potential or temperature, which induce a local fluid flow relative to the particle, and is tangential to the particle surface. 
In the limit of low Reynolds number (assuming negligible fluid and particle inertia), the solvent satisfies the Stokes equations,
\begin{equation}
    \bm{\nabla}\cdot\bm{v}=0
    \quad\text{and}\quad
    \bm{\nabla}\cdot\bm{\sigma}=\bm{0},
    \label{Stokes}
\end{equation}
where the constitutive equation for the stress is $\bm{\sigma}(\bm{x})=-p\bi{I}+\eta(\bm{\nabla}\bm{v}+(\bm{\nabla}\bm{v})^\text{T})$. Here, $p(\bm{x})$ is the pressure and $\eta$ is the shear viscosity. The particle satisfies Newton's equations, again in the absence of inertial effects, balancing the external force $\bm{F}^e$ and torque $\bm{L}^e$ acting on the particle by their hydrodynamic counterparts, i.e., 
\begin{equation}
    \bm{F}^e=-\int_{S}\bm{n}\cdot\bm{\sigma}\,\text{d}S \quad\text{and}\quad \bm{L}^e=-\int_{S}\bm{r}\times(\bm{n}\cdot\bm{\sigma})\,\text{d}S,
    \label{eq:newton}
\end{equation}
where $\bm{n}$ is the unit normal vector to the surface of the particle, directed into the fluid.

The reciprocal theorem relates the main flow $(\bm{v},\bm{\sigma})$ to an auxiliary model flow $(\hat{\bm{v}},\hat{\bm{\sigma}})$ according to \citep{lorentz1896eene}
\begin{equation}
\int_{S}\bm{n}\cdot\hat{\bm{\sigma}}\cdot\bm{v}\,\text{d}S
    = \int_{S}\bm{n}\cdot\bm{\sigma}\cdot\hat{\bm{v}}\,\text{d}S,
    \label{eq:rt}
\end{equation}
where $S$ refers to the particle surface only. This expression is valid for a particle in an unbounded fluid, as well as when the particle is confined either by rigid no-slip walls or by other planar, non-deformable boundaries, including stress-free surfaces and fluid-fluid interfaces, see Appendix \ref{sec:RT-interface}. As an auxiliary problem, we choose a model system with the same instantaneous configuration as the main problem that corresponds to the motion of a rigid, no-slip particle with translational and angular velocities, $\hat{\bm{V}}$ and $\hat{\bm{\Omega}}$, respectively. In the model problem an external force $\hat{\bm{F}}^e$ and torque $\hat{\bm{L}}^e$ act on the rigid particle. For convenience, we introduce the generalised velocity $\bi{V}=(\bm{V},\bm{\Omega})^{\text{T}}$ and the generalised external force $\bi{F}^e=(\bm{F}^e,\bm{L}^e)^{\text{T}}$ for the main problem. The corresponding generalised quantities for the auxiliary problem are $\hat{\bi{V}}=(\hat{\bm{V}},\hat{\bm{\Omega}})^{\text{T}}$ and $\hat{\bi{F}}^e=(\hat{\bm{F}}^e,\hat{\bm{L}}^e)^{\text{T}}$. Using the definitions \eqref{eq:slip} and \eqref{eq:newton} in \eqref{eq:rt} then yields
\begin{equation}
    \hat{\bi{F}}^e\cdot\bi{V}
    =\bi{F}^e\cdot\hat{\bi{V}} + \int_{S}\bm{n}\cdot\hat{\bm{\sigma}} \cdot\bm{v}_s \text{d} S .
    \label{eq:aux}
\end{equation}
In the auxiliary problem, we now use the linearity of Stokes equation to write $\hat{\bm{f}}=\bm{n}\cdot\hat{\bm{\sigma}}=\bm{\Pi}^{\text{T}}\cdot\hat{\bi{F}}^e$, where $\hat{\bm{f}}$ is the traction (force per unit area) on the surface of the rigid particle due to an external force $\hat{\bi{F}}^e$ acting on it. For reasons that will become apparent below, the rank-$2$ tensor field $\bm{\Pi}$ will henceforth be referred to as the grand propulsion tensor. Similarly, we write the generalised velocity of the rigid particle in the auxiliary problem as $\hat{\bi{V}}=\bi{M}\cdot\hat{\bi{F}}^e$, where $\bi{M}$ is the grand mobility tensor relating the force acting on the particle to its linear velocity. The resulting equation holds for arbitrary $\hat{\bi{F}}^e$ and so we obtain the generalised velocity of a confined active particle \citep{rallabandiMotionHydrodynamicallyInteracting2019}, 
\begin{equation}
    \bi{V} =  \bi{M}\cdot\bi{F}^e
    + \int_{S} \bm{\Pi}\cdot\bm{v}_s\,\text{d}S.
    \label{eq:main-result}
\end{equation}
The grand mobility and propulsion tensors depend only on the instantaneous configuration of the particle relative to nearby boundaries. If $\bi{M}$ and $\bm{\Pi}$ are known, equation \eqref{eq:main-result} is exact for a given slip velocity distribution. However, while the mobility is a well-known quantity that has been computed for many system geometries, it is not immediately clear how to proceed with the integral term containing the active contributions to the particle dynamics. While \eqref{eq:main-result} holds for an arbitrarily shaped particle, in the following we consider a spherical particle for simplicity.
The approximate \emph{many-body} grand propulsion tensor for spherical particles has been computed by \cite{rallabandiMotionHydrodynamicallyInteracting2019} for an unbounded fluid by using known results for linear flows.

Here, obviating the need for solving the Stokes equations in the bulk, we proceed by simultaneously expanding the surface fields, the slip $\bm{v}_s$ and the grand propulsion tensor $\bm{\Pi}$, directly at the surface of the particle, yielding \citep{singhManybodyMicrohydrodynamicsColloidal2015,turkFluctuatingHydrodynamicsAutophoretic2024}
\begin{equation}
    \bi{V}=
    \bi{M}\bm{\cdot}\bi{F}^e
    + \sum_{l=1}^\infty\sum_{\sigma\in\{s,a,t\}} \bm{\pi}^{(l\sigma)}\odot \bi{V}_s^{(l\sigma)},
    \label{eq:eom}
\end{equation}
where we refer to $\bm{\pi}^{(l\sigma)}$ as the propulsion tensors of the system and the $\bi{V}_s^{(l\sigma)}$ are the irreducible components of the slip. Here, $\sigma$ labels the symmetric ($\sigma=s$), anti-symmetric ($\sigma=a$), or trace ($\sigma=t$) part of the $l$th mode of the slip. The product $\odot$ implies a maximum contraction of indices. The details of this expansion are given in Appendix \ref{sec:explicit-theory}.

Using a Galerkin discretisation of the boundary integral representation of the Stokes equation, we have previously derived the mobility $\bi{M}$ and the propulsion tensors $\bm{\pi}^{(l\sigma)}$ in terms of derivatives of the Green's function of Stokes flow. The explicit expressions are provided in Appendix \ref{sec:propulsion-tensors}. The components of the slip $\bi{V}_s^{(l\sigma)}$ depend on the type of particle that is considered and in the case of self-phoretic particles have to be derived from the phoretic field, see Appendix \ref{sec:coupling}.  Equation \eqref{eq:eom} therefore describes the dynamics of an arbitrary spherical active particle in any system for which the Green's function of the Stokes equation is known.

By expressing the integral in \eqref{eq:main-result} as an infinite sum we have therefore established a direct connection between the previous results on active particles by \cite{rallabandiMotionHydrodynamicallyInteracting2019}, arrived at using the Lorentz reciprocal theorem, and \cite{singhManybodyMicrohydrodynamicsColloidal2015} and \cite{turkFluctuatingHydrodynamicsAutophoretic2024}, using a Galerkin discretisation of the surface fields.

\section{Autophoresis near a permeable surface}
\label{sec:autophoresis}
In this section we apply the result found in the previous section to the dynamics of an autophoretic particle near a plane boundary.
First, we define a particle that generates its own phoretic field, with gradients in said field driving a slip flow in a thin layer at the surface of the particle. We then define the boundary conditions that are imposed on the phoretic field and the flow field when the particle is located in the vicinity of a chemically permeable interface between two semi-infinite immiscible liquids.
Finally, we use this to characterise the long-time behaviours of a buoyant Janus particle near a variety of chemically and hydrodynamically non-trivial surfaces.

\subsection{Janus particle}
\label{sec:Janus}
\begin{figure}
    \centering
    \includegraphics[width=0.8\columnwidth]{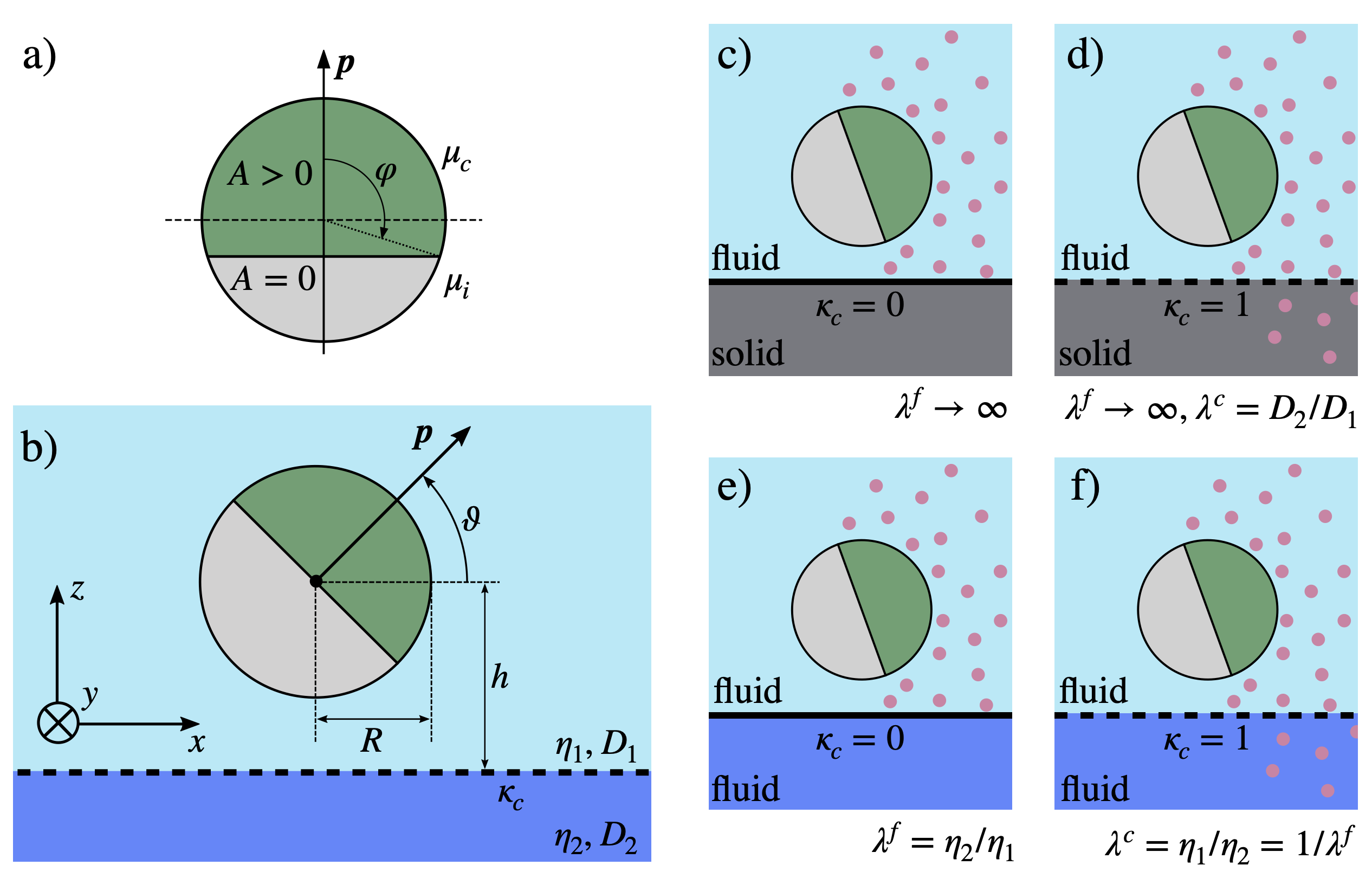}
    \vspace{-3mm}
    \caption{
        Janus particle and nearby planar surface schematics. Panel (a) shows a particle with an active cap ($A>0$) and an inert face ($A=0$) with phoretic mobilities $\mu_c$ and $\mu_i$, respectively. The cap size is determined by the contact angle $\varphi$. Panel (b) shows a half-covered ($\varphi=\pi/2$) particle of radius $R$ at a distance $h$ from a plane surface with a solute permeability $\kappa_c$ ($\kappa_c=0$ for an impermeable surface and $\kappa_c=1$ for a permeable surface). The interface is between two fluids of viscosities $\eta_1$ and $\eta_2$ and solute diffusivities $D_1$ and $D_2$ and lies in the $x$-$y$ plane. The assumed axisymmetry of the particle then allows us to limit its linear dynamics to the $x$-$z$ plane so that we can define the particle's orientation $\bm{p}$ via the angle $\vartheta$ to the surface. 
        Panels (c) -- (f) show a Janus particle near the four distinct types of boundaries considered here. The solute concentration produced by the particle is schematically shown as pink dots, diffusing into the particle's surroundings. In panels (c) and (d) a no-slip wall is represented by an interface with a diverging viscosity ratio $\lambda^f\rightarrow\infty$. Such a rigid surface can either be impermeable (panel (c), $\kappa_c=0$) or permeable (panel (d), $\kappa_c=1$) to the solutes. The latter is implied by the solute diffusing into the (porous) solid with non-zero solute diffusivity $D_2$. Panels (e) and (f) show an interface between two viscous liquids with a finite viscosity ratio $\lambda^f$. Again, this interface can either be impermeable (panel (e), $\kappa_c=0$) or permeable (panel (f), $\kappa_c=1$) to the solutes. For the latter, the Stokes-Einstein relation implies $D\propto 1/\eta$ so that the diffusivity ratio is given by the inverse of the viscosity ratio.
    }
    \label{fig:schematics}
\end{figure}

A Janus particle is an autophoretic particle that is characterised by a partial catalytic coating on its surface $S$ that generates a flux $A$ of solutes such that 
\begin{equation}
    -D\bm{n}\bm{\cdot}\bm{\nabla}c(\bm{x})=
    \begin{cases}
        A,\quad \bm{x}\in\text{ catalytic cap,}\\
        0,\quad \,\bm{x}\in\text{ inert face,}
    \end{cases}
    \label{eq:activity}
\end{equation}
where $c$ and $D$ are the local solute concentration and solute diffusivity, respectively. 
The quantity $\chi\coloneq -\cos\varphi$ parametrises the size of the catalytic cap, see figure \ref{fig:schematics}a.
In the limit of low P\'{e}clet number, in the bulk, the solutes diffuse freely according to the Laplace equation $\nabla^2c=0$. 
With the additional low Reynolds number assumption, see \eqref{Stokes}, the time dependence of both, the chemical and hydrodynamic governing equations, is suppressed, and all corresponding fields are quasi-steady.

The slip velocity distribution on the surface of the Janus particle can then be described by the phoretic boundary condition \citep{golestanian2007designing}
\begin{equation}
    \bm{v}_s(\bm{x})=\mu(\bm{x})\bm{\nabla}_s c(\bm{x})
    \quad\text{for}\quad \bm{x}\in S,
    \quad\text{where}\quad
    \mu(\bm{x})=
    \begin{cases}
        \mu_c,\quad \bm{x}\in\text{ catalytic cap,}\\
        \mu_i,\quad \,\bm{x}\in\text{ inert face,}
    \end{cases}
    \label{eq:phoretic-slip}
\end{equation}
where the gradient tangential to the particle surface is defined by $\bm{\nabla}_s=(\bi{I}-\bm{n}\bm{n})\bm{\cdot}\bm{\nabla}$. 
The phoretic mobility $\mu$ contains particle-solute interactions and varies between the inert face and the catalytic cap with a ratio $\beta=\mu_i/\mu_c$.
The boundary condition \eqref{eq:phoretic-slip} provides a means to derive the components of the slip $\bi{V}^{(l\sigma)}$ in the equations of motion \eqref{eq:eom} from the self-generated phoretic field. The explicit expressions are provided in Appendix \ref{sec:coupling}.

Janus particles in typical experiments are neither force- nor torque-free due to mismatches between particle and solvent densities and between the gravitational and geometric centres of the particle \citep{drescherDirectMeasurementFlow2010,ebbens2010pursuit,palacci2010sedimentation}. Gravity $\bm{g}$ therefore induces an external force and torque on the particle,
\refstepcounter{equation}
$$
    \bm{F}^e=m\bm{g},\qquad \bm{T}^e=r_m \bm{p}\times m\bm{g},
    \eqno{(\theequation{\mathit{a},\mathit{b}})}\label{eq:bottom-heaviness}
$$
where $m$ is the buoyant mass of the particle and $\bm{p}$ is its unit orientation vector. The distance between the particle's gravitational and geometric centres $r_m$ is a function of the cap size $\varphi$ and the relative thickness of the catalytic cap, as well as the density ratio between the cap- and particle materials, and is given in Appendix \ref{sec:geometric-cap-model}.

\subsection{Permeable surface}
\label{sec:confined-dynamics}
We now consider the specific system in which the Janus particle is confined to the positive half-space $z>0$ by an infinite surface in the $x$-$y$ plane, see figure \ref{fig:schematics}b. In addition to the boundary condition on the particle surface \eqref{eq:activity}, the solute obeys 
\refstepcounter{equation}
$$
    c^{(2)} = \kappa_c\, c^{(1)},\quad 
    D_1\,\partial_z c^{(1)} = D_2\,\partial_z c^{(2)}\;\,
    \text{ for }\;z=0\quad 
    \text{ and }\quad c^{(i)}\rightarrow 0\;\text{ for }\; r\rightarrow\infty,
    \eqno{(\theequation{\mathit{a}\text{-}\mathit{c})}}
    \label{eq:c-bc}
$$
where $c^{(i)}$ and $D_i$ with $i=1,2$ are the concentration field and solute diffusivity in the regions $z>0$ and $z<0$, respectively. The solute permeability $\kappa_c\in\{0,1\}$ indicates whether the surface is impermeable ($\kappa_c=0$) or permeable ($\kappa_c=1$) to the solutes. The particle is assumed to be the only source of solutes so that the solute concentration vanishes far from the particle. The Green's function of the Laplace equation satisfying these boundary conditions is given in Appendix \ref{sec:explicit-theory}.

Furthermore, we assume that the surface is the planar, non-deformable boundary between two semi-infinite immiscible liquids. Using the same notation as above for the regions above and below the surface, in addition to the phoretic slip boundary condition \eqref{eq:phoretic-slip} the fluid flow and stress satisfy the conditions~\citep{blakeNoteImageSystem1971},
\refstepcounter{equation}
$$
    v^{(1)}_\rho = v^{(2)}_\rho, \quad 
    v^{(1)}_z = v^{(2)}_z=0, \quad
    \sigma^{(1)}_{\rho z} = \sigma^{(2)}_{\rho z}\quad\text{for }\; z=0\quad \text{ and }\quad
    \bm{v}\rightarrow\bm{0}\;\text{ for }r\rightarrow \infty,
    \eqno{(\theequation{\mathit{a}\text{-}\mathit{d}})}
    \label{eq:v-bc}
$$
where the index $\rho=x,y$ lies in the plane of the interface. Therefore, across the interface the fluid is characterised by continuous tangential velocity, vanishing normal velocity, and continuous tangential stress. In the absence of a background flow the fluid is at rest far away from the particle. The Green's function of the Stokes equation satisfying these boundary conditions is given in Appendix \ref{sec:explicit-theory}.

The boundary conditions (\ref{eq:c-bc}) and (\ref{eq:v-bc}) allow us to to characterise the plane surface by its viscosity ratio and, in case it is permeable to the solutes, by its solute diffusivity ratio,
\refstepcounter{equation}
$$
    \lambda^f=\frac{\eta_2}{\eta_1},\qquad \kappa_c=
    \begin{cases}
        1\quad\rightarrow\quad \lambda^c=D_2/D_1,\\
        0.
    \end{cases}
    \eqno{(\theequation{\mathit{a},\mathit{b}})}\label{eq:ratios}
$$
With this we define four distinct types of boundaries considered here, see panels (c) - (f) in figure \ref{fig:schematics}. A no-slip wall can be represented by an interface with a diverging viscosity ratio $\lambda^f\rightarrow\infty$. Such a rigid surface can either be impermeable ($\kappa_c=0$) or permeable ($\kappa_c=1$) to the solutes. In the latter case the solid can be interpreted as porous with a non-zero solute diffusivity $D_2$. A free surface or fluid-gas interface ($\lambda^f=0$) can be defined analogously. For the case of a finite viscosity ratio $\lambda^f$ the surface represents an interface between two viscous liquids, which again can either be impermeable ($\kappa_c=0$) or permeable ($\kappa_c=1$) to the solutes. In the latter case, since for a viscous liquid the Stokes-Einstein relation implies $D\propto 1/\eta$, the viscosity and solute diffusivity ratios are related such that $\lambda^c=1/\lambda^f$. It is worth noting that $\lambda^f=1$ does not correspond to an unbounded fluid, because the normal component of the fluid velocity at the non-deformable interface still vanishes according to the boundary conditions \eqref{eq:v-bc}. 

Assuming that gravity points towards the surface so that $\bm{g}=-g\hat{\bm{z}}$, we introduce a scale for the particle's cap-heaviness, $G_A = mg/6\pi\eta RU$, measuring the strength of gravitational effects relative to the particle's activity. Here, $1/6\pi\eta R$ is the translational mobility of a spherical particle of radius $R$ in an unbounded fluid and $U=\mu_cA/D$ is the typical speed of a self-phoretic particle. Since $m$ is the buoyant mass of the particle, $G_A>0$ ($G_A<0$) implies that gravity pushes (pulls) the particle towards (away from) the surface with a gravitational torque turning the catalytic cap towards (away from) the surface.

In this system the dynamics of an axisymmetric Janus particle with a given cap size $\chi$ and cap-heaviness $G_A$ are fully parametrised by its (relative) height $H\coloneq h/R$ above the surface and its orientation $\vartheta$, see figure \ref{fig:schematics}b. Its relative lateral position is denoted by $X\coloneq x/R$. The resulting dynamical system is obtained directly from \eqref{eq:eom} for each bounding surface and consists of the coupled equations $(\dot{H}(H,\vartheta),\,\dot{\vartheta}(H,\vartheta))$ and the independent lateral dynamics $\dot{X}(H,\vartheta)$, where a dotted variable implies a derivative with respect to time $t$. In this description of the dynamical system we have left the following system parameters implicit for brevity: the particle's catalytic cap-size $\chi$, the cap-heaviness $G_A$, the chemical permeability of the nearby surface $\kappa_c$, the solute diffusivity contrast $\lambda^c$, and the viscosity contrast $\lambda^f$. 

In the following, we include only long-ranged chemo-hydrodynamic effects, neglecting linear and angular interactions that decay faster than $H^{-3}$ and $H^{-4}$, respectively. We prevent particle-boundary contact with a short-ranged repulsive potential (Appendix \ref{sec:repulsive-surface}). 
While such regularisations are standard, they can affect particle behaviour by altering the dynamical system so that one dynamical state is favoured over another or by introducing spurious oscillations \citep{ibrahimHowWallsAffect2016,lintuvuoriHydrodynamicOscillationsVariable2016,bayatiDynamicsPlanarWalls2019,shumHydrodynamicInteractionsSedimenting2025}.

\subsection{Long-time behaviours}
\label{sec:results}

\begin{figure}
    \centering
    \includegraphics[width=\columnwidth]{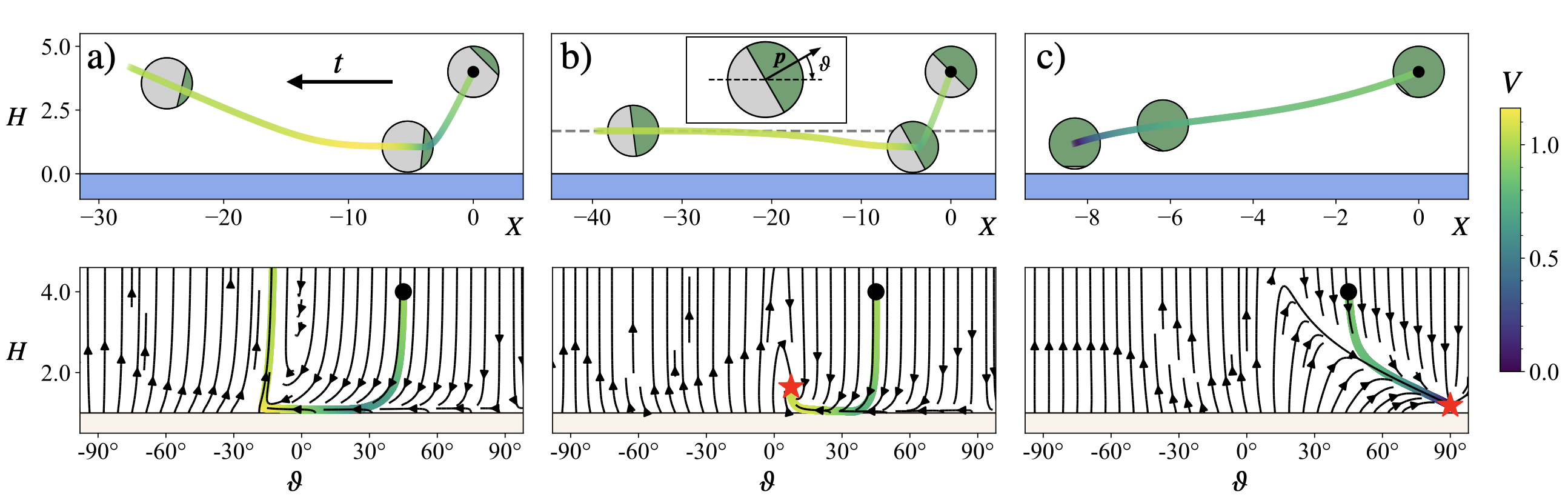}
    \vspace{-6mm}
    \caption{
    Examples of typical long-time behaviours of a neutrally buoyant Janus particle ($\beta=0.9$) near a permeable fluid-fluid interface ($\kappa_c=1$ and $\lambda^f=1/\lambda^c=10$) as a function of its catalytic cap size. 
    The initial position and orientation in each case are  $X=0$, $H=4$ and $\vartheta=45^\circ$, with the particles moving from right to left in the top row as indicated by the direction of time $t$ in panel (a).
    The inset in the top row of panel (b) defines the particle orientation vector $\bm{p}$ and the associated angle $\vartheta$ to the plane of the interface. 
    The normalised speed $V$ of the particle along its real- and phase-space trajectories is indicated by the corresponding colour bar.
    In the real-space trajectories in the top row, the particle size and orientation are not shown to scale with respect to the $x$-axis. The particle's initial position is marked by a black dot and the second fluid in the region $z\leq0$ is indicated in blue. 
    In the corresponding phase plots in the bottom row, the initial condition (black dot), the phase-space trajectory and any fixed points (red star) are shown. The area $z\leq1$ cannot physically be reached by the particle. 
    Panel (a) shows a particle with a small cap ($\chi=-0.6$) escaping the interface by being chemo-hydrodynamically reflected by it.
    In panel (b) a half-covered particle ($\chi=0$) settles to a steady skating state at a fixed height (indicated by a dashed grey line) and tilt angle.
    In panel (c) a particle with a very large cap ($\chi=0.9$) enters a stable hovering state, effectively acting as a stationary micro-pump for the surrounding fluid. 
    Movie 1, movie 2 and movie 3 of the online supplementary material show the trajectories in panels (a), (b) and (c), respectively.
    }
    \label{fig:trajectories}
\end{figure}

\begin{figure}
    \centering
    \includegraphics[width=\columnwidth]{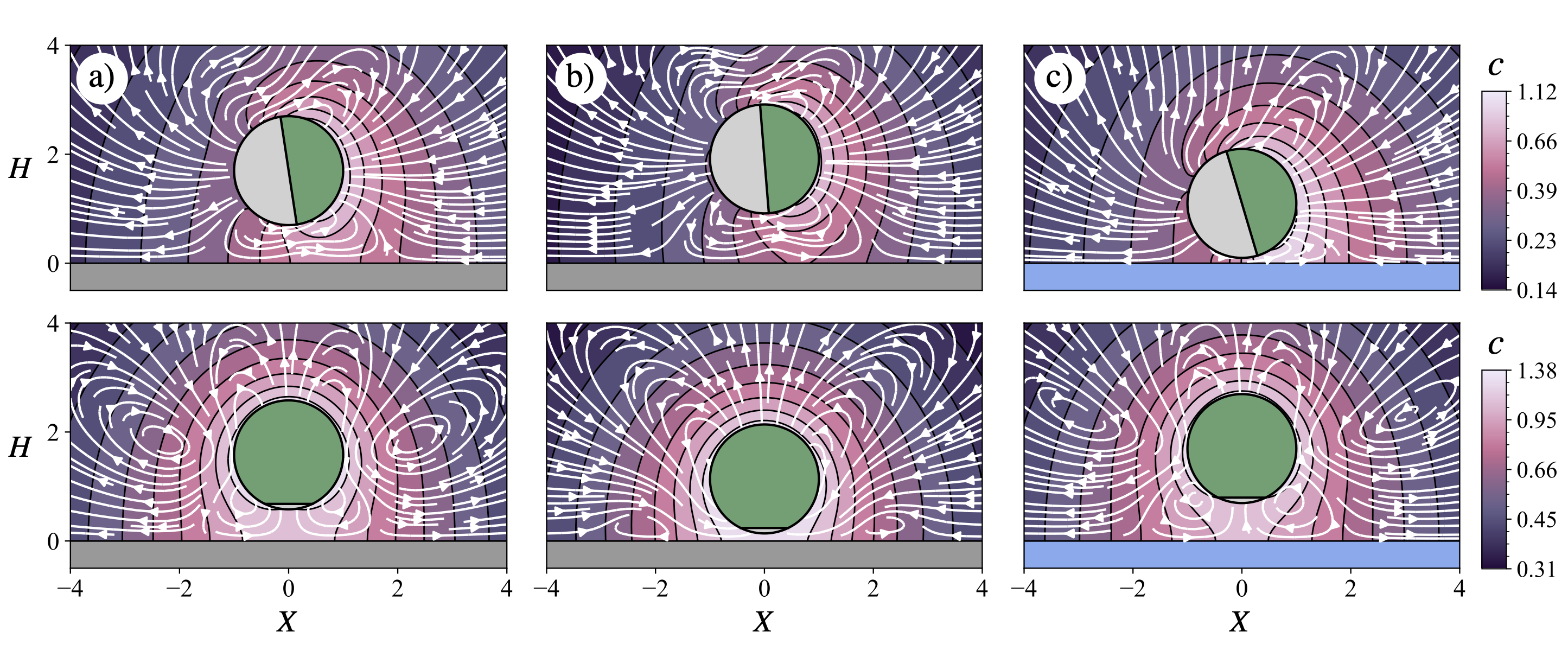}
    \vspace{-6mm}
    \caption{
    Examples of the generated solute concentration (contours with overlaid pseudo-colour map on a log-scale) and flow field (direction indicated by white arrows) of neutrally buoyant Janus particles ($\beta=0.9$) in a stable skating (top, where $\chi=0$) or hovering (bottom, where $\chi=0.9$) state near various surfaces. The panels correspond to the following surfaces: (a) impermeable wall ($\kappa_c=0$ and $\lambda^f\rightarrow\infty$), (b) permeable wall ($\kappa_c=1$, $\lambda^c=0.1$ and $\lambda^f\rightarrow\infty$), and (c) impermeable fluid-fluid interface ($\kappa_c=0$ and $\lambda^f=1$). The corresponding skating angles are $8.6^\circ$, $4.6^\circ$ and $16.5^\circ$, respectively. It is worth noting that, for an impermeable surface, the contour lines meet the boundary at a right angle and the corresponding vector field ($\bm{\nabla}c$) becomes purely tangential to this `no-flux' boundary. 
    }
    \label{fig:conc-flow}
\end{figure}

\begin{figure}
    \centering
    \vspace{-5mm}
    \includegraphics[width=\columnwidth]{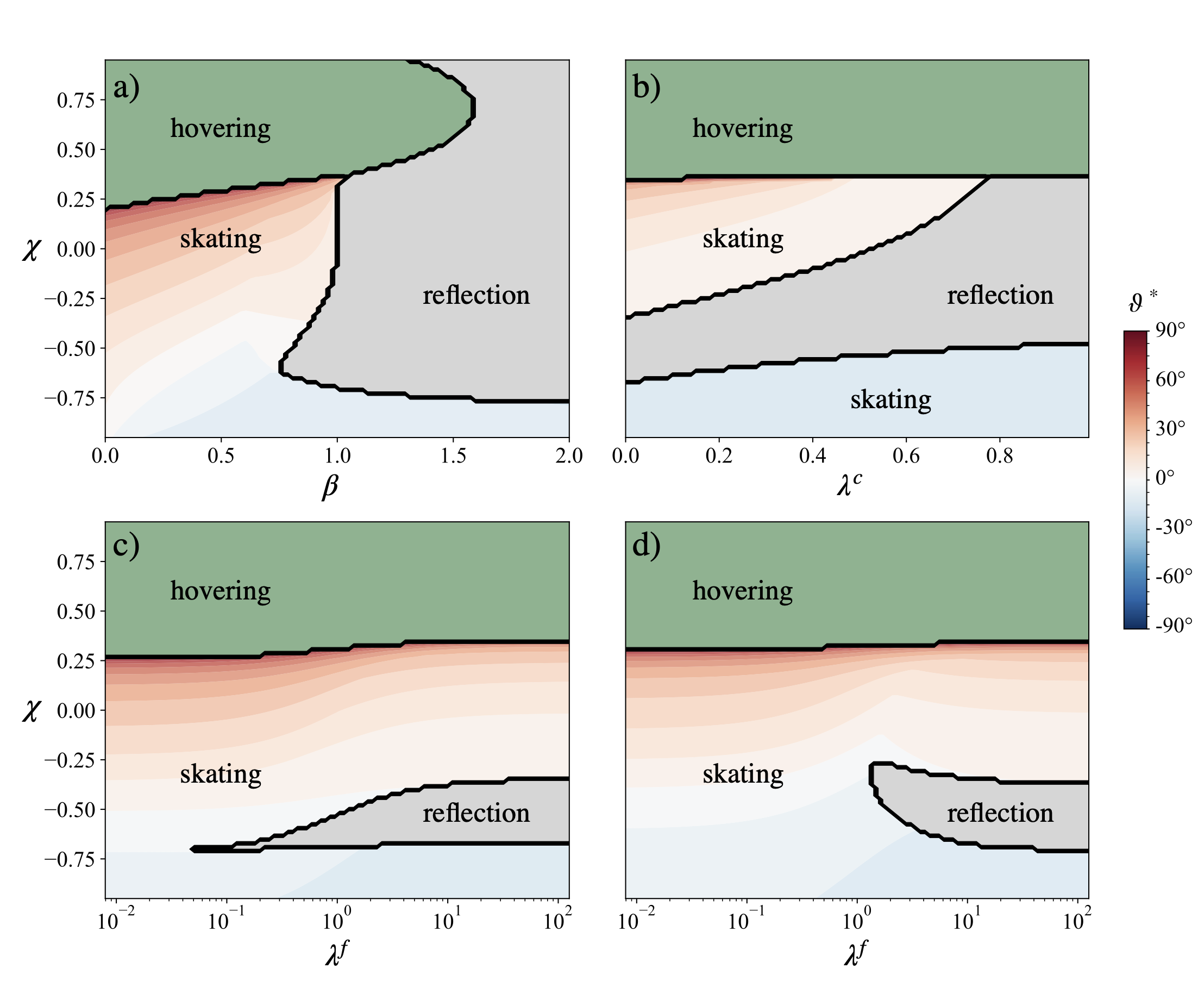}
    \vspace{-7mm}
    \caption{
        Long-time behaviours of a neutrally buoyant Janus particle near a variety of surfaces with initial conditions $H_0=2$ and $\vartheta_0=45^\circ$ as a function of its cap size $\chi$ (where $|\chi|\leq0.95$) and other parameters. 
        The particle is deemed to have escaped the wall if $H>30$ at any time. 
        For the skating state, the steady tilt angle $\vartheta^*$ is indicated by the colour bar.
        Panel (a) shows the phase diagram of a particle near an impermeable rigid wall ($\kappa_c=0$ and $\lambda^f\rightarrow\infty$) as a function of the size of its  phoretic mobility ratio $\beta$.
        For panels (b-d) we set $\beta=0.9$. 
        Panel (b) shows the phase diagram near a permeable rigid wall ($\kappa_c=1$, finite $\lambda^c$ and $\lambda^f\rightarrow\infty$) as a function of the diffusivity ratio $\lambda^c$. 
        Panel (c) shows the phase diagram near an impermeable fluid-fluid interface ($\kappa_c=0$ and finite $\lambda^f$) as a function of the viscosity ratio $\lambda^f$.
        Panel (d) shows the phase diagram near a permeable fluid-fluid interface ($\kappa_c=1$ and finite $\lambda^c=1/\lambda^f$) as a function of the viscosity ratio $\lambda^f$.
    }
    \label{fig:pd-all}
\end{figure}

\begin{table}
  \begin{center}
  \setlength{\tabcolsep}{10pt} % Adjust column spacing
\def~{\hphantom{0}}
  \begin{tabular}{p{0.25\textwidth}p{0.22\textwidth}p{0.36\textwidth}}
         & \textbf{Bounding surface} & \textbf{Method}\\[5pt]
         \cite{uspalSelfpropulsionCatalyticallyActive2015},\newline
         \cite{bayatiDynamicsPlanarWalls2019},\newline
         \cite{dasFloorCeilingSlidingChemically2020} &
         Wall &
         Collocation method (BEM)\newline
         (numerical)
         \\[25pt]
         \cite{ibrahimDynamicsSelfphoreticJanus2015,ibrahimHowWallsAffect2016} &
         Wall &
         Multipole method \newline
         (analytical)
         
         \\[15pt]
         \cite{simmchenTopographicalPathwaysGuide2016},\newline
         \cite{uspalActiveJanusColloids2019} &
         Wall with phoretic slip &
         Collocation method (BEM) \newline
         (numerical)
         \\[15pt]
        \cite{mozaffariSelfdiffusiophoreticColloidalPropulsion2016,mozaffariSelfpropelledColloidalParticle2018} &
         Wall &
         Bispherical coordinates \newline
         (semi-analytical)
         \\[15pt]

         This paper &
         Wall, \newline
         Permeable wall, \newline
         Fluid-fluid interface, \newline
         Permeable interface &
         Galerkin method \newline
         (analytical)
  \end{tabular}
  \caption{This paper in the context of previous theoretical work on the steady states of Janus particles near planar surfaces. Several of these contributions have also taken into account the effect of gravity. 
  Each method listed offers distinct advantages and limitations.
  Collocation methods, while accurate in the near field and highly versatile, incur a high computational cost. 
  Semi-analytical approaches based on bispherical coordinates achieve comparable accuracy at much lower cost, but are restricted to simple geometries. 
  Multipole and Galerkin methods, being analytical, provide the greatest physical insight, yet rely on far-field expansions, limiting their validity to sufficiently large particle-boundary separations.
  }
  \label{tab:overview}
  \end{center}
\end{table}

In this section, we investigate the long-time behaviours of a Janus particle near various bounding surfaces as described earlier, examples of which are illustrated in figure \ref{fig:trajectories}. 
We categorise these long-time behaviours into three primary states: (1) chemo-hydrodynamic reflection, where the particle is repelled from the boundary, escaping its influence; (2) skating, a state in which the particle moves steadily at a fixed height and constant tilt angle $\vartheta^*$; and (3) hovering, a stationary state characterised by the particle remaining fixed at an angle $\vartheta^*=90^\circ$, effectively acting as a microscopic fluid pump.
In terms of the underlying dynamical system, these states correspond to (1) no fixed point; (2) a stable fixed point for which $\dot{H}=\dot{\vartheta}=0$, but $\dot{X}\neq0$; and (3) a stable fixed point satisfying $\dot{H}=\dot{\vartheta}=\dot{X}=0$.
Additionally, under certain gravitational conditions, more complex oscillatory states can arise, as discussed below.

In figure \ref{fig:conc-flow} we show the solute concentration and flow fields (accurate to order $r^{-3}$ in the distance from the particle) generated by a Janus particle in a steady state near various interfaces -- rigid, permeable and fluid fluid.
We illustrate how the interfacial properties impact both, the typical skating and hovering states of a half-covered particle and a particle with a large catalytic cap, respectively.
We systematically investigate these effects in the following.

First, we note that the phase planes shown in figure \ref{fig:trajectories} are characteristic of two-timescale dynamics, where the fast variable $H$ quasi-instantaneously adjusts to the slow variable $\vartheta$ \citep{uspalSelfpropulsionCatalyticallyActive2015}. 
The emerging quasi-equilibrium curve $\dot{H}=0$, on which lie both the skating and hovering states, can be understood by considering the leading order dynamics in the particle's vertical motion. 
The equations of motion \eqref{eq:eom} for a neutrally buoyant ($G_A=0$) particle with a uniform phoretic mobility distribution on its surface ($\beta=1$) yield the  equilibrium condition:
\begin{equation}
    \tfrac{1}{4}(1-\chi)^2\sin\vartheta 
    =\frac{1+\chi}{2304}
    \left[
        288\,\Lambda^c
        +5(1-\chi)\chi\,\Lambda^f (1-3\sin^2\vartheta)
    \right]
    H^{-2}
    +\mathcal{O}(H^{-3}),
    \label{eq:leading-order}
\end{equation}
where $\Lambda^c=(1-\kappa_c\lambda^c)/(1+\kappa_c\lambda^c)$ such that $\Lambda^c\leq1$ in general, and $\Lambda^c=1$ for a chemically impermeable surface.
For brevity, we have also introduced $\Lambda^f=(2+3\lambda^f)/(1+\lambda^f)$.
The details of the calculation leading to this result can be found in Appendix \ref{sec:explicit-theory}.
The left-hand side of \eqref{eq:leading-order} is the $z$-component of the particle's velocity in an unbounded fluid. 
The right-hand side comprises the contributions to the particle's vertical velocity due to interactions with the nearby surface. 
To leading order in the inverse distance between the particle and the surface, the origin of these interactions is two-fold.
The first term arises from the chemical monopole -- the total flux of solutes emanating from the particle -- interacting with the chemically permeable surface. 
The second term is the leading-order hydrodynamic interaction and arises from the anisotropic distribution of catalyst on the surface of the particle. It is worth noting that this term vanishes for a half-covered particle, for which $\chi=0$. In this case, hydrodynamic interactions only emerge at $\mathcal{O}\left(H^{-3}\right)$ unless there is a non-uniform distribution of surface mobility ($\beta\neq1$). 

From \eqref{eq:leading-order} we can deduce that for small angles $H\propto \vartheta^{-1/2}$. Therefore, in the skating state, a steeper tilt angle $\vartheta^*$ leads to a reduced equilibrium distance $H^*$ from the surface, as illustrated in the top row of panel (c) of figure \ref{fig:conc-flow}.

We can also use equation \eqref{eq:leading-order} to estimate the equilibrium height at which a particle with a large catalytic cap may hover ($\vartheta=90^\circ$) above a surface as a function of the surface's chemo-hydrodynamic properties. 
For high catalytic coverages, we expect chemical effects to dominate the hydrodynamic interactions between the particle and the surface. Thus, neglecting the second term, we obtain $H^*=\sqrt{\Lambda^c/[2(1-\chi)]}$ as a leading order estimate for the hovering height. 
This simple result allows us to make two predictions. 
On the one hand, we expect particles of larger cap-sizes $\chi$ to hover at a larger distance to the surface. 
On the other hand, chemical permeability (so that $\Lambda^c<1$) of the surface is expected to reduce this distance, as illustrated in the bottom row of panel (b) of figure \ref{fig:conc-flow}. 
It is worth noting that our leading order estimate for the hovering height is in slight disagreement with a previous result that has been obtained for a chemically impermeable wall ($\Lambda^c=1$) by \citet{uspalSelfpropulsionCatalyticallyActive2015}, whose calculations yield a different pre-factor. However, the first term in \eqref{eq:leading-order}, which leads to our estimate, matches a result obtained by \citet{yarivWallinducedSelfdiffusiophoresisActive2016} for the wall-induced motion of an isotropically active particle ($\chi=1$) in the far-field. 

A direct analysis of the steady tilt angle in the skating state or of the dynamics for particles with non-uniform phoretic mobility distributions, i.e., for $\beta\neq1$, is more involved as more and higher order terms have to be taken into account. 
Therefore, by numerically integrating the equations of motion \eqref{eq:eom}, we generate phase diagrams depicting the distinct long-time behaviours discussed above as a function of both, particle and surface properties. This is illustrated in figure \ref{fig:pd-all}.
The influence of gravity on particles that are cap-heavy and not perfectly density-matched with the surrounding solution is discussed separately, with corresponding results presented in figure \ref{fig:pd-gravity}. 
For reference and comparison, previous theoretical work on confined Janus particles is summarised in table \ref{tab:overview}.

We set $\mu_c>0$ and $\beta\geq0$ so that the particle is chemo-repulsive, i.e., the particle behaves as an inert-side forward swimmer.  
The dynamics of the particle depend on its initial orientation $\vartheta_0=\vartheta(t=0)$ and relative height $H_0=H(t=0)$ above the surface. 
For orientations initially directed away from the surface ($\vartheta_0<0$), the particle typically escapes the boundary, as demonstrated in Appendix \ref{sec:more-pds}.
To explore states bound to the surface, we use the initial conditions $\vartheta_0=45^\circ$ and $H_0=2$ in the following.
Although variations in the initial angle $\vartheta_0>0$ can shift phase boundaries slightly, they do not yield qualitatively different states (Appendix \ref{sec:more-pds}).

\subsubsection{Impermeable rigid wall}
\label{sec:wall}
We first revisit the well-studied scenario of a Janus particle near an impermeable rigid wall ($\kappa_c=0$, $\lambda^f\rightarrow\infty$). 
Typical results for the  particle’s long-term dynamics as a function of its catalytic coverage $\chi$ and phoretic mobility ratio $\beta$ are illustrated in figure \ref{fig:pd-all}a.
An interesting region of the phase diagram is the very narrow window of skating for a uniform phoretic mobility distribution on the particle's surface ($\beta=1$) for large cap-sizes ($\chi>0$). 
From this narrow region, the skating phase expands instantly for $\beta<1$. 
This is because $\beta\neq1$ introduces an additional, purely chemical mechanism for rotating the particle, absent in the case of uniform phoretic mobility (Appendix \ref{sec:coupling}).
On the other hand, for $\beta>1$ the particle is largely reflected by the wall.  

A previously unreported observation is that in our far-field approximation particles with small catalytic caps ($\chi\lesssim-0.75$) can skate at slightly negative equilibrium tilt angles $\vartheta^*$.
This is the case across all surface properties shown in figure \ref{fig:pd-all}.
We find, however, that this region of the phase diagram is sensitive to the choice of repulsive potential between the particle and the wall. While we choose to impose a stiff, short-ranged potential (range$/R=1.1$, see Appendix \ref{sec:repulsive-surface}), emulating a hard-core repulsion between the swimmer and the surface, \citet{ibrahimHowWallsAffect2016}, who used a similar truncation of chemo-hydrodynamic effects as is used here, employ a longer-ranged, flatter potential. This increased range of the potential can remove a stable skating or hovering state, with the particle eventually escaping the boundary. It is worth noting that other regions of the phase diagram are found to be robust against such changes in the repulsive potential. 
However, since other studies \citep{uspalSelfpropulsionCatalyticallyActive2015,mozaffariSelfdiffusiophoreticColloidalPropulsion2016} -- which used high‑fidelity numerical methods -- also failed to report any small‑cap skating and instead observed small‑capped particles being reflected by a wall, we must assume that this phenomenon is an artefact of the truncated dynamics used in our simulations.

The hard-core repulsion potential employed here is designed such that skating and hovering heights are independent of the potential's influence, unless these heights are precisely within the potential's operative range. Two distinct scenarios emerge: (a) without the mitigating effect of the short-range repulsive potential, the particle's trajectory would result in a collision with the boundary, or (b) long-range chemo-hydrodynamic interactions are inherently sufficient to prevent the particle from contacting the boundary. 
In the former situation, when the particle is within the range of the potential while hovering, we find that this state is stabilised purely because of the potential, such that is hovers at a height roughly equal to the range of the stiff potential. 
In the latter case, however, since the particle's closest approach to the wall is beyond the short-ranged potential's reach, the potential plays no role in the particle's dynamics.
For a direct comparison of the long-term behaviours of a Janus particle near various surfaces with and without the addition of a repulsive potential, see Appendix \ref{sec:repulsive-surface}.

 \subsubsection{Permeable rigid wall}
 \label{sec:wall-perm}
Next, we examine the effect of chemical permeability in a rigid wall ($\kappa_c=1$, $\lambda^c=D_2/D_1$). 
Figure \ref{fig:pd-all}b illustrates how permeability influences a neutrally buoyant Janus particle's long-time dynamics (with $\beta=0.9$), as a function of its catalytic coverage $\chi$ and the diffusivity ratio of the permeable surface $\lambda^c$. 
For $\lambda^c=0$, the surface effectively behaves as impermeable. 
Increasing the diffusivity ratio reduces the equilibrium tilt angle during skating for a fixed cap size. 
Qualitatively, this can be explained as follows.
As is shown in Appendix \ref{sec:angular-velocity}, chemical interactions with the surface tend to orient the catalytic cap away from it, whereas hydrodynamic interactions turn the cap towards it. Hence, permeable surfaces, by reducing chemical interactions, result in shallower skating angles, potentially leading to particles escaping the surface above a certain critical value of $\lambda^c$.

 \subsubsection{Impermeable fluid-fluid interface}
 \label{sec:intf}
We now explore neutrally buoyant Janus particles (with $\beta=0.9$) near a chemically impermeable interface between two immiscible fluids ($\kappa_c=0$). 
The corresponding phase diagram (figure \ref{fig:pd-all}c) illustrates how, given a cap size $\chi$, equilibrium skating angles $\vartheta^*$ steepen with decreasing viscosity ratios $\lambda^f$. 
At finite viscosity ratios, hydrodynamic interactions with the interface are reduced compared to a rigid wall, resulting in larger equilibrium tilt angles for skating states; see Appendix \ref{sec:angular-velocity} for an illustration.  
Consequently, particles skating along rigid walls might attain a stationary hovering state when near a fluid-fluid interface. 
Analogously, a particle that is reflected by a rigid wall might skate along a fluid-fluid interface.
This analysis holds even for the limiting case of a stress-free surface ($\lambda^f=0$).

 \subsubsection{Permeable fluid-fluid interface}
 \label{sec:intf-perm}
Investigating a neutrally buoyant Janus particle (with $\beta=0.9$) near a chemically permeable fluid-fluid interface ($\kappa_c=1$, $\lambda^c=1/\lambda^f$), we illustrate the set of identified long-time behaviours in figure \ref{fig:pd-all}d. 
Remarkably, for an interface with a fluid of comparatively small viscosity, so that  $\lambda^f<1$ ($\lambda^c>1$), the bounding surface becomes chemically attractive to the catalytic cap.
This leads to qualitatively different long-time dynamics than for the same particle near an impermeable interface (see figure \ref{fig:pd-all}c) and could be exploited experimentally as an indirect measure of interfacial permeability.
For shallow initial orientations of the particle and $\lambda^f\ll1$ we briefly discuss the emergence of an `inverted hovering' state in Appendix \ref{sec:more-pds}. 
As in the hovering state, here the particle effectively acts as a stationary micro-pump for the fluid, but now with its catalytic cap facing the boundary ($\vartheta=-90^\circ$) instead of facing away from it. 
This distinctive behaviour is absent near impermeable surfaces.

\begin{figure}
    \centering
    \includegraphics[width=\columnwidth]{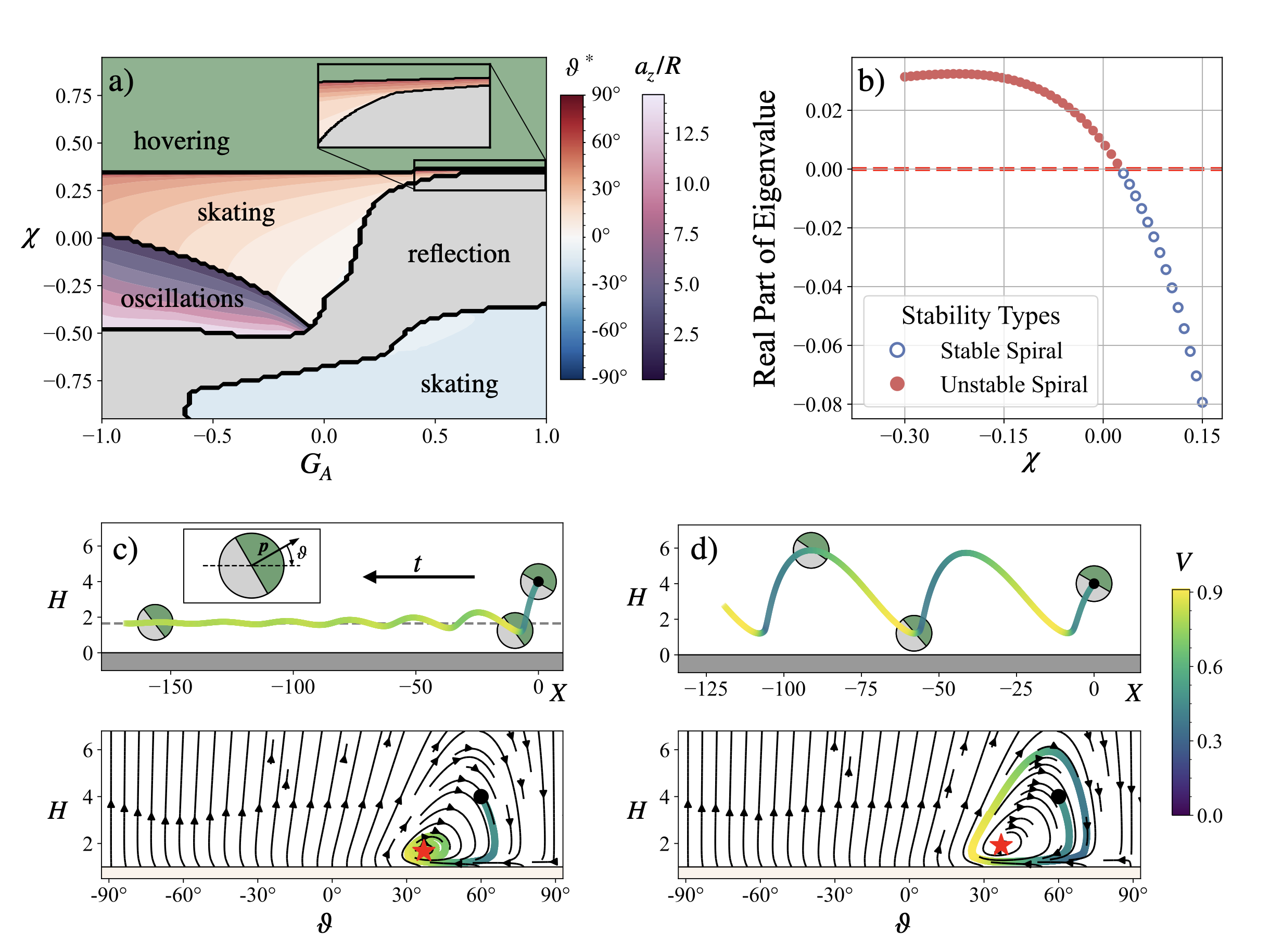}
    \vspace{-6mm}
    \caption{
        Long-time behaviours of a buoyant Janus particle with $\beta=0.9$ near an impermeable rigid wall ($\kappa_c=0$ and $\lambda^f\rightarrow\infty$) under the influence of gravity. 
        The particle's initial conditions are $H_0=2$ and $\vartheta_0=45^\circ$. 
        Panel (a) shows the particle's phase behaviour as a function of its cap size $\chi$ (where $|\chi|\leq0.95$) and its cap-heaviness $G_A$.
        The particle is deemed to have escaped the wall if $H>30$ at any time.
        For the skating and oscillating states, the skating angle $\vartheta^*$ and the relative amplitude of the oscillations in $z$-direction $a_z/R$ are indicated by the respective colour bars.
        The inset shows the detailed dynamics for $0.33<G_A<1$ and $0.26<\chi<0.38$.
        In panels (b) to (d) we set $G_A=-1$.
        Panel (b) shows the results of a linear stability analysis around the fixed point of the dynamical system described in the main text as a function of the cap-size $\chi$. 
        The real part of the complex-conjugated eigenvalues of the $2\times2$ Jacobian matrix at the fixed point is shown to cross zero at $\chi\approx0.027$, suggesting the occurrence of a Hopf bifurcation, where for smaller cap-sizes a periodic solution arises. 
        The eigenvalues are always complex-valued, indicating stable (inwards) or unstable (outwards) spiralling dynamics near the fixed point. 
        Panels (c) and (d) show sample real-space (top) and phase-plane trajectories (bottom) before ($\chi=0.06$) and after ($\chi=-0.05$) the Hopf bifurcation, respectively, illustrating the transition from a stable to an unstable spiral with an emerging limit cycle in the phase plane with decreasing cap-size. 
        In panels (c) and (d) the initial position and orientation (indicated by a black dot) of the particle are $X=0$, $H=4$ and $\vartheta=60^\circ$, with the particle moving from right to left in real-space as indicated by the direction of time $t$.
        The inset defines the particle orientation vector $\bm{p}$ and the associated angle $\vartheta$ to the wall. 
        The normalised speed $V$ of the particle along its real- and phase-space trajectories is indicated by the corresponding colour bar.
        In real space, the rigid solid in the region $z\leq0$ is indicated in grey.
        In phase space, the region $z\leq1$ cannot physically be reached by the particle.
        Movie 4 and movie 5 of the online supplementary material show the trajectories in panels (c) and (d), respectively. 
    }
    \label{fig:pd-gravity}
\end{figure}

\subsubsection{Gravitational effects}
\label{sec:gravity}
Finally, we study gravitational effects on a Janus particle (with $\beta=0.9$) near an impermeable rigid wall \citep{dasFloorCeilingSlidingChemically2020,mozaffariSelfdiffusiophoreticColloidalPropulsion2016}, as shown in figure \ref{fig:pd-gravity}. 
Considering the experimentally relevant case of a silica sphere with a platinum cap (buoyant density ratio $K\approx 17$) with a relative cap thickness $d/R\approx 5\cdot 10^{-3}$ \citep{dasFloorCeilingSlidingChemically2020}, we find the emergence of a stable oscillating state for positively buoyant particles ($G_A<0$), not previously observed. 

A linear stability analysis around the skating fixed point $(H^*,\vartheta^*)$ of the dynamical system $(\dot{H}(H,\vartheta),\,\dot{\vartheta}(H,\vartheta))$ shows that as $G_A$ and/or $\chi$ are varied, a supercritical Hopf bifurcation may occur, indicating the emergence of a periodic solution and thus, a transition from transient to stable oscillatory behaviour, see panels (b) to (d) in figure \ref{fig:pd-gravity}. 
Physically, the emerging limit cycle in the phase plane describes the particle initially swimming towards the surface, where chemo-hydrodynamic interactions turn its catalytic cap towards the wall, eventually detaching the particle from the surface (reflection). As the interactions with the wall weaken with distance, gravitational effects become dominant, pulling the cap away from the surface, so that the particle once again swims towards the wall until this cycle repeats itself.

While here we only consider moderate values $|G_A|\leq1$, stronger gravitational effects with $|G_A|\gg1$ can lead to additional dynamics. 
A state similar to the `inverted hovering' reported in section \ref{sec:intf-perm} for a chemically permeable fluid-fluid interface, has been found by \citet{mozaffariSelfdiffusiophoreticColloidalPropulsion2016} to be induced by gravity near an impermeable rigid wall.
For Janus particles in a container with both a floor and a ceiling, \citet{dasFloorCeilingSlidingChemically2020} have identified regions of parameter space in which sliding states may emerge simultaneously at both bounding surfaces.

\section{Discussion}
\label{sec:discussion}

In this paper, we have developed a theoretical framework for the dynamics of autophoretic particles near chemically and hydrodynamically complex surfaces. By leveraging the Lorentz reciprocal theorem and employing a Galerkin discretisation of the relevant surface fields, we derived a governing equation for the motion of a confined active particle. Obviating the need to solve the bulk fluid equations explicitly, this method is computationally efficient and so particularly well-suited for exploring a wide range of particle properties and boundary conditions. 
We applied this framework to study the dynamics of Janus particles near rigid, permeable, and fluid-fluid interfaces, identifying and categorising a set of stable long-time behaviours, including surface-bound skating along surfaces, stationary hovering states, and chemo-hydrodynamic reflection by the boundary. 
As with other analytical approaches, our framework is based on a truncated far-field expansion. This restriction limits its quantitative accuracy in the near-field regime, where higher-order terms or fully numerical methods are required. Nevertheless, we have demonstrated that our results qualitatively agree with the existing literature, including studies employing semi-analytical expansions to very high order or high-fidelity numerical simulations.

Our results highlight the intricate interplay between chemical and hydrodynamic interactions in determining the particle's long-time behaviours with respect to nearby surfaces. For example, we find that solute permeability of the boundary significantly alters the particle's orientation and motility. 
When the particle is near a chemically impermeable rigid wall, its catalytic cap is chemically repelled from and hydrodynamically attracted to the wall. 
Under certain circumstances, this can lead to a stable skating state with a specific steady tilt angle. 
For a permeable wall, an increase in the diffusivity contrast (less chemical repulsion of the catalytic cap) results in shallower skating angles and, in some cases, enables the particle to escape the wall. 
Similarly, at fluid-fluid interfaces, we observe a strong dependence of the dynamics on the viscosity ratio between the two fluids. 
When the viscosity of the adjacent fluid is comparatively small, hydrodynamic interactions with the interface are diminished, leading to the emergence of stable hovering states that for the same type of particle may be absent near rigid boundaries.
Notably, in cases where the interface is both chemically permeable and viscosity-stratified, we identify an inverted hovering state, in which the catalytic cap of the stationary particle faces the interface, a feature that does not appear in impermeable systems of neutrally buoyant particles. 
Reduced chemical and hydrodynamic interactions between the particle and a permeable fluid-fluid interface may account for some of the effects observed by \citep{palaciosGuidanceActiveParticles2019} for the guidance of a Janus particle along an oil-water interface. 
Their observations include a reduction in propulsion speed compared with the bulk and a seemingly increased significance of Brownian fluctuations when compared with the dynamics near a rigid wall, leading to reduced particle retention times   \citep{mozaffariSelfpropelledColloidalParticle2018}. 
While in this study we have neglected the effect of thermal fluctuations for simplicity, they can be expected to play a role in the dynamics of Janus particles. Importantly, owing to the linearity of Stokes flow, contributions from Brownian motion can be appended to the deterministic equations of motion. A systematic framework for including such stochastic contributions in the dynamics of autophoretic particles has been presented by \citet{turkFluctuatingHydrodynamicsAutophoretic2024}.

From a methodological standpoint, our approach complements existing computational techniques for studying the dynamics of autophoretic particles near bounding surfaces; see table \ref{tab:overview} for an overview. Boundary element methods (BEM) provide highly accurate predictions, particularly at small particle-surface separations, but are computationally expensive and require careful meshing. Methods based on multipole expansions and bispherical coordinates offer efficient analytical and semi-analytical solutions, respectively, but often suffer limited applicability to complex geometries and restricted flexibility in capturing diverse boundary conditions. Our Galerkin-based approach, while inevitably suffering from truncation errors in the near-field regime, provides a computationally efficient alternative that allows for systematic exploration of broad parameter spaces. By capturing the essential interactions governing particle dynamics, our method can serve as a first step in identifying motility regimes and informing more computationally intensive methods, such as BEM, for detailed analysis of specific parameter sets.

In summary, in this paper we provide new insights into the behaviour of autophoretic particles in complex environments. By systematically accounting for both hydrodynamic and chemical interactions in an analytical framework, we have mapped out the stable motility states of such particles near a range of boundaries. Future work is necessary to extend our approach to account for additional effects such as thermal fluctuations, Marangoni forces, and external flow and phoretic fields, further bridging the gap between theoretical predictions and experimental observations. Moreover, the computational efficiency of our method makes it particularly well-suited for exploring many-body interactions in suspensions of active particles. Extending the framework to study collective behaviours, such as emergent clustering and dynamic self-organization near complex interfaces, could provide valuable insights into both synthetic and biological micro-swimmer systems.

\backsection[Acknowledgements]{We thank W.E. Uspal and Y.-N. Young for helpful discussions.}

\backsection[Funding]{
G.T. and H.A.S. thank the Princeton Center for Complex Materials, a MRSEC (NSF DMR-2011750), for support of this research.
R.S. acknowledges support from the Indian Institute of Technology, Madras, India and their seed and initiation grants as well as a Start-up Research Grant, SERB, India (SERB file number: SRG/2022/000682)
}

\backsection[Declaration of Interests]{The authors report no conflict of interest.}

\newpage
\appendix

\section{Reciprocal theorem for a particle confined by a planar, non-deformable boundary}
\label{sec:RT-interface}
In this section we show that the reciprocal theorem for a particle that is confined by a planar, non-deformable boundary between two semi-infinite immiscible liquids simplifies to equation \eqref{eq:rt}. In figure \ref{fig:schematics}b we show a schematic of this. The velocity field for the fluid in which the particle resides is denoted by  $(\bm{v}^{(1)},\bm{\sigma}^{(1)})$ and the velocity field in the confining fluid is $(\bm{v}^{(2)},\bm{\sigma}^{(2)})$. Hatted variables are used for the auxiliary problem. Applying the reciprocal theorem in fluid $1$ leads to  
\begin{equation}
    \sum_S\int_{S}\bm{n}\cdot\hat{\bm{\sigma}}^{(1)}\cdot\bm{v}^{(1)}\,{\rm d} S
    =
    \sum_S\int_{S}\bm{n}\cdot\bm{\sigma}^{(1)}\cdot\hat{\bm{v}}^{(1)}\,{\rm d} S,
    \label{eq:RT-fluid1}
\end{equation}
where the sum is over both, the surface of the particle $S_p$ and the interface $S_i$. The normal vector to the interface is directed into fluid $1$.
In fluid $2$, the reciprocal theorem yields
\begin{equation}
    \int_{S_i}\bm{n}\cdot\hat{\bm{\sigma}}^{(2)}\cdot\bm{v}^{(2)}\,{\rm d} S
    =
    \int_{S_i}\bm{n}\cdot\bm{\sigma}^{(2)}\cdot\hat{\bm{v}}^{(2)}\,{\rm d} S.
    \label{eq:RT-fluid2}
\end{equation}
Using the boundary conditions \eqref{eq:v-bc} in \eqref{eq:RT-fluid2} and subtracting this from \eqref{eq:RT-fluid1} yields the desired result, where we dropped the superscript $(1)$ indicating fluid $1$ in \eqref{eq:rt}. It is worth noting that for either, a rigid no-slip wall, or a negligibly deforming stress-free surface, equation \eqref{eq:rt} is trivially satisfied.

\section{Explicit particle dynamics}
\label{sec:explicit-theory}
In this section we provide explicit expressions for the mobility and propulsion tensors and the components of the slip in the equations of motion. First, we explicitly state the expansions of the surface fields, leading from the integral in \eqref{eq:main-result}, containing the active contributions to the particle dynamics, to the infinite sum containing only known quantities in \eqref{eq:eom}.
We then give the mobilities and propulsion tensors we have previously derived in terms of derivatives of the Green's function of the Stokes equation. 
Finally, we provide the relevant slip modes arising for a Janus particle, given its chemical activity and phoretic mobility distributions. 

For convenience, in the absence of other, i.e., non-gravitational, external forces and torques, in the following we rescale forces and torques by $mg$ and $mgR$, respectively. Mobilities are rescaled by $1/6\pi\eta R$, the  translational mobility of a spherical particle of radius $R$ in an unbounded fluid. Furthermore, we rescale lengths by $R$, concentrations by $RA/D$, velocities by $U=\mu_cA/D$, angular velocities by $U/R$ and pressures by $\eta U/R$ and rename the thus non-dimensionalised variables so that they read the same. With this, the scale for the particle's cap-heaviness introduced in the main text, $G_A=mg/6\pi\eta RU$, emerges naturally in the equations of motion \eqref{eq:eom}.

\subsection{Expansion of the surface fields}
\label{sec:expansion}
We first introduce the so-called tensor spherical harmonics (TSH) \citep{hess2015tensors}:
\begin{equation}
    \bi{Y}^{(l)}(\bm{n})=(2l-1)!!\;\bm{\Delta}^{(l)}\odot\bi{n}^{(l)}.
    \label{eq:tsh}
\end{equation}
Here, $\bm{\Delta}^{(l)}$ is a rank-$2l$ tensor that projects a tensor of rank-$l$ onto its symmetric and traceless part, the product $\odot$ implies a maximum contraction of indices and $n^{(l)}_{\alpha_1\dots\alpha_l}\coloneq n_{\alpha_1}\dots n_{\alpha_l}$ with $\bm{n}$ the unit normal vector to the surface of a sphere pointing into the fluid. The components of the symmetric and traceless tensor $\bi{Y}^{(l)}$ of rank-$l$, expressed in terms of the polar angles, are isomorphic to the spherical harmonics $Y_l^m$. The TSH form an irreducible basis on the surface of a sphere of radius $R$ with the orthogonality relation
\begin{equation}
    \int_S Y^{(l)}_{\alpha_1\dots\alpha_l}
    \,Y^{(l')}_{\gamma_1\dots\gamma_{l^\prime}}
    \,\mathup{d}S
    =\delta_{ll'}\frac{1}{w_l\tilde{w}_l}
    \,\Delta^{(l)}_{\alpha_1\dots\alpha_l,{\gamma_1\dots\gamma_l}},
    \label{eq:tsh-ortho}
\end{equation}
where by convention $w_l=1/l!(2l-1)!!$ and $\tilde{w}_l=(2l+1)/4\pi R^2$. 

We proceed by simultaneously expanding the surface fields, the slip $\bm{v}_s$ and the grand propulsion tensor $\bm{\Pi}$, in this basis. For the slip we obtain \citep{turkStokesTractionActive2022} 
\begin{multline}
    \bm{v}_s=\sum_{l=1}^{\infty}w_{l-1}\bi{V}_s^{(l)}\odot\bi{Y}^{(l-1)}(\bm{n}),
    \quad\text{where} \\
    \bi{V}_s^{(l)}=\bm{\Delta}^{(l)}\odot\bi{V}_s^{(ls)} 
    -\tfrac{l-1}{l}\bm{\Delta}^{(l-1)}\odot\big(\bm{\epsilon}\cdot\bi{V}_s^{(la)}\big)
    + \tfrac{2l-3}{2l-1}\bm{\Delta}^{(l-1)}\odot\big(\bi{I}\cdot\bi{V}_s^{(lt)}\big).
    \label{eq:slip-expansion}
\end{multline}
Here, $\bm{\epsilon}$ is the Levi-Civita tensor and $\bi{I}$ is the identity matrix.
The expansion coefficients for the slip, $\bi{V}_s^{(l)}$, are rank-$l$ tensors that, by construction are symmetric and traceless in their last $(l-1)$ indices. The components $\bi{V}_s^{(l\sigma)}$ are symmetric and traceless tensors of rank $l$ for $\sigma=s$ (symmetric part of $\bi{V}_s^{(l)}$), $l-1$ for $\sigma=a$ (anti-symmetric part of $\bi{V}_s^{(l)}$) and $l-2$ for $\sigma=t$ (trace part of $\bi{V}_s^{(l)}$). Similarly, the expansion of the grand propulsion tensor yields
\begin{equation}
    \bm{\Pi}=\sum_{l=1}^\infty \tilde{w}_{l-1}\, \bm{\pi}^{(l)}\odot\bi{Y}^{(l-1)}(\bm{n}),
    \label{eq:propulsion-expansion}
\end{equation}
where we refer to $\bm{\pi}^{(l)}$ as the propulsion tensors of the system. They are tensors of rank-$(l+1)$ that are symmetric and traceless in their last $(l-1)$ indices. Using the expansions of the slip \eqref{eq:slip-expansion} and the grand propulsion tensor \eqref{eq:propulsion-expansion}, together with the orthogonality relation for TSHs  \eqref{eq:tsh-ortho} in equation \eqref{eq:main-result}, the integral describing the active contributions to the particle dynamics in the latter is transformed to the infinite sum in \eqref{eq:eom}. In the latter we have taken into account that the symmetric and traceless components $\bi{V}_s^{(l\sigma)}$ of the slip coefficients impose their symmetries on the propulsion tensors. To make the translational (superscript $T$) and rotational (superscript $R$) components of the propulsion tensors explicit, we can write  $\bm{\pi}^{(l\sigma)}=(\bm{\pi}^{(T,l\sigma)},\bm{\pi}^{(R,l\sigma)})^\text{T}$. The structure of the problem then implies that $\bm{\pi}^{(R,l\sigma)}=\tfrac{1}{2}\bm{\nabla}\times\bm{\pi}^{(T,l\sigma)}$.

\subsection{Mobilities and propulsion tensors}
\label{sec:propulsion-tensors}
We have previously derived the mobilities and propulsion tensors in equation \eqref{eq:eom} in terms of derivatives of the Green's function $\bi{G}$ of the Stokes equation. For an arbitrary system we can write the Green's function as the sum \citep{smoluchowski1911mutual}
\begin{equation}
    \bi{G}(\bm{R}_1,\bm{R}_2)=\bi{G}^o(\bm{r})+\bi{G}^*(\bm{R}_1,\bm{R}_2),
    \label{eq:green-stokes}
\end{equation}
where $\bi{G}^o$ is the Oseen tensor for unbounded Stokes flow \citep{oseenHydrodynamik1927}:
\begin{equation}
    \bi{G}^o(\bm{r})=\tfrac{1}{8\pi r}(\bi{I}+\hat{\bm{r}}\hat{\bm{r}}),
    \label{eq:Oseen}
\end{equation}
with $\hat{\bm{r}}=\bm{r}/r$, where $r=|\bm{r}|$ and $\bm{r}=\bm{R}_1-\bm{R}_2$. The position vectors $\bm{R}_1$ and $\bm{R}_2$ indicate the field and source points, respectively. The term $\bi{G}^*$ is the correction necessary to satisfy additional boundary conditions in the system. 

The dimensionless grand mobility tensor can be written as 
\begin{equation}
    \bi{M}=
    \begin{pmatrix}
        \bi{M}^{TT} & \bi{M}^{TR}\\
        \bi{M}^{RT} & \bi{M}^{RR}
    \end{pmatrix},
    \label{eq:grand-mobility}
\end{equation}
where purely translational, purely rotational and mixed terms are made explicit. For example, for a particle near a plane boundary by symmetry the mobilities depend only on the relative distance $H$ between the centre of the particle and the bounding surface. Keeping terms up to $\mathcal{O}\left (H^{-3}\right )$ the following terms are included in our analysis:
\begin{align}
    \bi{M}^{TT}&\approx \bi{I} + 6\pi\big(1+\tfrac{1}{6}\nabla^2_1 + \tfrac{1}{6}\nabla^2_2\big)\bi{G}^*, & 
    \bi{M}^{TR} &\approx 3\pi\,\bm{\nabla}_2\times\bi{G}^*, \nonumber\\[5pt]
    \bi{M}^{RR} &\approx \tfrac{3}{4}\bi{I} + \tfrac{3\pi}{2}\,\bm{\nabla}_1\times\bm{\nabla}_2\times\bi{G}^*, &
    \label{eq:mobilities}
\end{align}
where we have introduced the short-hand notation $\bm{\nabla}_i=\bm{\nabla}_{\bm{R}_i}$.

For the modes $l\sigma\in\{1s,2a\}$ corresponding to rigid-body motion we find $\bm{\pi}^{(1s)}=(-\bi{I},\bi{0})^{\rm tr}$ and $2\bm{\pi}^{(2a)}=(\bi{0},-\bi{I})^{\rm tr}$. For an unbounded fluid, these are the only non-zero coefficients. For a particle near a plane boundary, all other propulsion tensors decay as  $\bm{\pi}^{(T,l\sigma)}\sim H^{-l}$.  Keeping terms up to $\mathcal{O}\left (H^{-3}\right )$ we obtain \citep{singhManybodyMicrohydrodynamicsColloidal2015,turkFluctuatingHydrodynamicsAutophoretic2024}
\begin{align}
    \bm{\pi}^{(T,2s)}\colon\bi{V}_{s}^{(2s)}
    &\approx \tfrac{20\pi}{3}\bm{\nabla}_2\bi{G}^{*}\colon\bi{V}_{s}^{(2s)}, &
    \bm{\pi}^{(T,3s)}\vdots\bi{V}_{s}^{(3s)}
    &\approx \tfrac{7\pi}{6} \bm{\nabla}_2\bm{\nabla}_2\bi{G}^{*}\,\vdots\,\bi{V}_{s}^{(3s)},\nonumber\\[5pt]
    \bm{\pi}^{(T,3a)}\colon\mathsfbi{V}_{s}^{(3a)}
    &\approx \tfrac{4\pi}{9}\bm{\nabla}_2\big(\bm{\nabla}_2\times\mathsfbi{G}^{*}\big)\colon\mathsfbi{V}_{s}^{(3a)}, &
    \bm{\pi}^{(T,3t)}\cdot\bi{V}_{s}^{(3t)}
    &\approx -\tfrac{2\pi}{5} \nabla^{2}_2\bi{G}^{*}\cdot\bi{V}_{s}^{(3t)}.
\end{align}
Here, multiple vertically arranged dots imply a contraction of multiple Cartesian indices. The correction to the Green's function corresponding to the boundary conditions \eqref{eq:v-bc} is given by \citep{blakeNoteImageSystem1971}
\begin{equation}
    \bi{G}^*(\bm{R}_1,\bm{R}_2)=
    \bm{\mathcal{M}}^f\cdot\bi{G}^o(\bm{r}^*) 
    - \tfrac{\lambda^f}{1+\lambda^f}\Big(2z_2\bm{\mathcal{M}}\cdot\bm{\nabla}^*\bi{G}^o(\bm{r}^*)\cdot \hat{\bm{z}}
    -z_2^2\nabla^{*2}\bi{G}^o(\bm{r}^*)\cdot\bm{\mathcal{M}}\Big),
    \label{eq:green-corr-v}
\end{equation}
where $\bm{r}^*=\bm{R}_1-\bm{\mathcal{M}}\cdot\bm{R}_2$, involving the mirroring operator $\mathcal{M}_{\alpha\beta}=\delta_{\alpha\rho}\delta_{\beta\rho}-\delta_{\alpha z}\delta_{\beta z}$ with the index $\rho=x,y$ in the plane of the surface.
Additionally, we define $\bm{\nabla}^*\coloneq \bm{\nabla}_{\bm{r}^*}$, the matrix $\mathcal{M}^f_{\alpha\beta}=\tfrac{1-\lambda^f}{1+\lambda^f}\delta_{\alpha\rho}\delta_{\beta\rho}-\delta_{\alpha z}\delta_{\beta z}$, and $z_2$ is the $z$-component of the source point $\bm{R}_2$. In the region $z<0$, the required Green's function is
\begin{equation}
    \bi{G}^{\,z<0}(\bm{R}_1,\bm{R}_2)= \tfrac{1}{1+\lambda^f}
        \left[ 
            2 \bi{G}^o(\bm{r}) \cdot(\bi{I} - \hat{\bm{z}}\hat{\bm{z}})
            - 2 z_2 \bm{\nabla}\bi{G}^o(\bm{r})\cdot \hat{\bm{z}}
            - z_2^2 \nabla^2 \bi{G}^o(\bm{r})
        \right],
    \label{eq:green-lower-v}
\end{equation}
where $\bm{\nabla}\coloneq \bm{\nabla}_{\bm{r}}$.

\subsection{Autophoretic slip}
\label{sec:autophoretic-theory-explicit}
The below results are obtained by solving the boundary integral representation of the Laplace equation, given the boundary conditions in equation \eqref{eq:c-bc} in an irreducible basis of tensor spherical harmonics (TSH), defined in equation \eqref{eq:tsh}. The aim is to obtain the phoretic slip for a Janus particle, given its chemical activity and phoretic mobility distributions. 
Given the particle's unit orientation vector $\bm{p}$ and the unit normal vector to its surface pointing into the fluid $\bm{n}$, we expand the particle's activity in a basis of TSH:
\begin{equation}
    A(\chi,\,\bm{p}\cdot\bm{n})=\sum_{q=0}^\infty \bi{A}^{(q)}(\chi,\,\bm{p})\odot\bi{Y}^{(q)}(\bm{n}).
    \label{eq:activity-expansion}
\end{equation}
Here, $\chi=-\cos\varphi$ parametrises the size of the catalytic cap, where $\varphi$ is the contact angle of the cap. 
For an axisymmetric particle we can write $\bi{A}^{(q)}(\chi,\,\bm{p})=A_q(\chi)\bi{Y}^{(q)}(\bm{p})$. The activity mode $A_q$ ($q\geq1$) contributes a fluid flow of order $r^{-q}$ and $r^{-(q+2)}$, where $r$ is the distance to the flow disturbance. Long-ranged flows decaying no faster than $r^{-3}$, therefore, are captured by the first four modes $q\leq3$. The normalised activity distribution \eqref{eq:activity} can then be approximated by setting \citep{ibrahimHowWallsAffect2016}
\begin{align}
    A_0&=\tfrac{1}{2}(1+\chi),& A_1&=\tfrac{3}{4}(1-\chi^2),\nonumber\\
    A_2&=-\tfrac{5}{24}(1-\chi^2)\chi, & A_3&=-\tfrac{7}{1440}(1-\chi^2)(1-5\chi^2).
    \label{eq:activity-coeffs}
\end{align}

\begin{figure}
    \centering
    \includegraphics[width=.8\columnwidth]{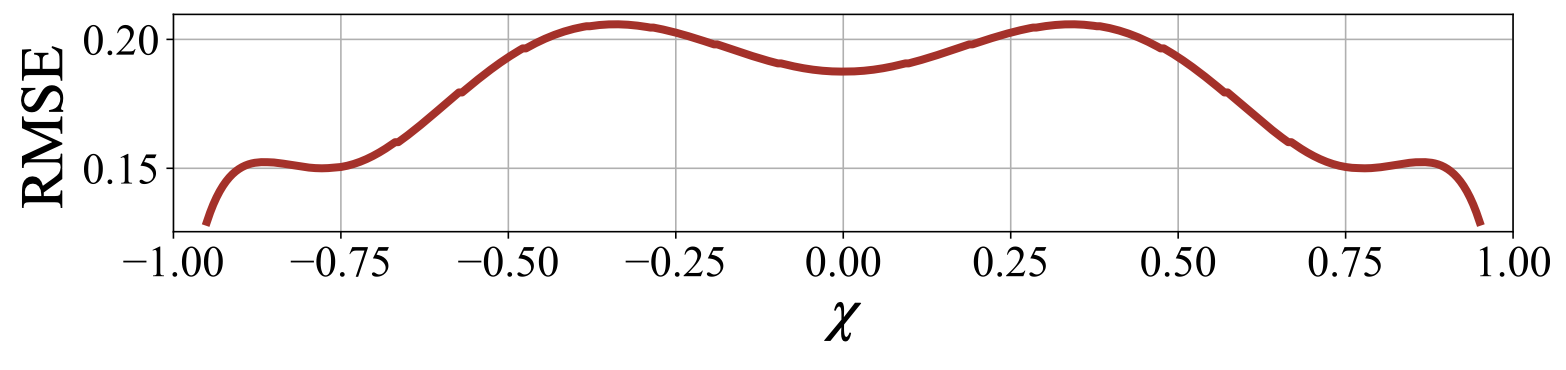}
    \vspace{-3mm}
    \caption{
        The continuous Root Mean Square Error (RMSE) as a measure for the cumulative error in the approximation of the surface activity for a Janus particle as a function of its cap size for values $-0.95\leq\chi\leq 0.95$. 
    }
    \label{fig:activity_error}
\end{figure}

In order to quantify the quality of this approximation, we temporarily label the activity distribution in \eqref{eq:activity} as $A^{\rm exact}(\chi, \bm{p}\cdot\bm{n})$ and the approximated activity distribution in \eqref{eq:activity-expansion} with the coefficients \eqref{eq:activity-coeffs} as $A^{\rm approx}(\chi, \bm{p}\cdot\bm{n})$. First, we define the local error of the approximation as 
\begin{equation}
    \Delta A(\chi, \bm{p}\cdot\bm{n}) = A^{\rm approx}(\chi, \bm{p}\cdot\bm{n}) - A^{\rm exact}(\chi, \bm{p}\cdot\bm{n})
\end{equation}
With this, we can then define the continuous Root Mean Square Error (RMSE) as a measure for the cumulative error for each cap size $\chi$:
\begin{equation}
    {\rm RMSE}(\chi) = \sqrt{\frac{1}{2}\int_{-1}^{1}\Delta A(\chi, \bm{p}\cdot\bm{n})^2 \,{\rm d}(\bm{p}\cdot\bm{n})}.
\end{equation}
This can be evaluated numerically and is shown in figure \ref{fig:activity_error} for values $-0.95\leq\chi\leq 0.95$. Notably, the cumulative error in the approximation of the surface activity is smaller for very small and very large caps. 

\subsection{Surface concentration}
\label{sec:chem-problem}
Expanding the solute concentration at the surface of the particle as in equation \eqref{eq:activity-expansion} with coefficients $\bi{C}^{(q)}$ the boundary integral representation of the Laplace equation yields \citep{singhCompetingChemicalHydrodynamic2019,turkFluctuatingHydrodynamicsAutophoretic2024}
\begin{equation}
    \bi{C}^{(q)}= \sum_{q'} \bm{\mathcal{E}}^{(q,q')}\odot\bi{A}^{(q')},
    \label{eq:elastances}
\end{equation}
where $\bm{\mathcal{E}}^{(q,q')}$ is a tensor of rank $q+q'$ which can be written in terms of derivatives of the Green's function $L$ of the Laplace equation. For an arbitrary system we can write the Green's function as the sum
\begin{equation}
    L(\bm{R}_1,\bm{R}_2)=L^o(r)+L^*(\bm{R}_1,\bm{R}_2),
    \label{eq:green-conc}
\end{equation}
with $r=|\bm{r}|$ and $\bm{r}=\bm{R}_1-\bm{R}_2$, where $\bm{R}_1$ and $\bm{R}_2$ are the field and source points, respectively. Here, $L^o(r)=1/4\pi r$ is the Green's function in an unbounded domain and $L^*$ is the correction necessary to satisfy additional boundary conditions in the system. For an unbounded system, the tensor diagonalises with diagonal elements given by $\mathcal{E}_q=1/(q+1)$, while for a permeable interface defined by the boundary conditions \eqref{eq:c-bc} we have previously derived a general expression. By symmetry, its entries only depend on the relative distance $H$ between the centre of the particle and the interface and off-diagonal elements decay as $H^{q+q'+1}$. Retaining terms up to $\mathcal{O}\left (H^{-4}\right )$ yields
\begin{align}
    &\bm{\mathcal{E}}^{(1,0)}A^{(0)}\approx6\pi\bm{\nabla}_1 L^*A^{(0)},&
    &\bm{\mathcal{E}}^{(1,1)}\cdot\bm{A}^{(1)}\approx\tfrac{1}{2}(\bi{I}+6\pi\bm{\nabla}_1\bm{\nabla}_2 L^*)\cdot\bm{A}^{(1)},\nonumber\\
    &\bm{\mathcal{E}}^{(1,2)}\colon\bi{A}^{(2)}\approx 2\pi \bm{\nabla}_1 \bm{\nabla}_2 \bm{\nabla}_2 L^* \colon\bi{A}^{(2)},&
    &\bm{\mathcal{E}}^{(2,0)}A^{(0)}\approx\tfrac{10\pi}{9}\bm{\nabla}_1\bm{\nabla}_1 L^*A^{(0)},\nonumber\\
    &\bm{\mathcal{E}}^{(2,1)}\cdot\bm{A}^{(1)}\approx\tfrac{10\pi}{3}\bm{\nabla}_1\bm{\nabla}_1\bm{\nabla}_2 L^* \cdot\bm{A}^{(1)},&
    &\bm{\mathcal{E}}^{(2,2)}\colon\bi{A}^{(2)}\approx\tfrac{1}{3}\bi{A}^{(2)},&
    \nonumber\\
    &\bm{\mathcal{E}}^{(3,0)}A^{(0)}\approx 7\pi \bm{\nabla}_1\bm{\nabla}_1\bm{\nabla}_1 L^* A^{(0)},&
    &\bm{\mathcal{E}}^{(3,3)}\,\vdots\,\bi{A}^{(3)}\approx\tfrac{1}{4}\bi{A}^{(3)}.
    \label{eq:elastances}
\end{align}
Once again, we have used the short-hand notation $\bm{\nabla}_i=\bm{\nabla}_{\bm{R}_i}$. 
The correction to the Green's function of the Laplace equation corresponding to the boundary condition \eqref{eq:c-bc} is given by
\begin{equation}
    L^*(\bm{R}_1,\bm{R}_2)=\tfrac{1-\kappa_c\lambda^c}{1+\kappa_c\lambda^c}L^o(r^*),
    \label{eq:green-corr-c}
\end{equation}
where $r^*=|\bm{r}^*|$ and the vector $\bm{r}^*$ is defined below equation \eqref{eq:green-corr-v}.
In the region $z<0$, the required Green's function is
\begin{equation}
    L^{z<0}(\bm{R}_1,\bm{R}_2)= \tfrac{2\kappa_c}{1+\kappa_c\lambda^c}L^o(r).
    \label{eq:green-lower-c}
\end{equation}

\subsection{Phoretic slip}
\label{sec:coupling}
The distribution of surface concentration drives a phoretic slip via $\bm{v}_s=\mu\bm{\nabla}_sc$, see equation \eqref{eq:phoretic-slip}. We expand the phoretic mobility, which varies between the catalytic cap ($\mu_c$) and the inert face ($\mu_i$) with a ratio $\beta=\mu_i/\mu_c$, analogously to the activity \eqref{eq:activity-expansion}, keeping the coefficients $\bi{M}^{(q)}(\bm{p})$ for $q\leq3$, with the normalised mode strengths
\begin{align}
    M_0&=\beta+\tfrac{1}{2}(1-\beta)(1+\chi),& M_1&=\tfrac{3}{4}(1-\beta)(1-\chi^2),\nonumber\\
    M_2&=-\tfrac{5}{24}(1-\beta)(1-\chi^2)\chi, & M_3&=-\tfrac{7}{1440}(1-\beta)(1-\chi^2)(1-5\chi^2).
    \label{eq:mobility-coeffs}
\end{align}
We combine this with the expansion of the phoretic slip in equation \eqref{eq:slip-expansion} and note that, in an unbounded fluid, coefficients labelled by $l$ generate a fluid flow decaying as $r^{-l}$. 

Finally, we can write the phoretic slip boundary condition in an irreducible basis,
\begin{equation}
    \bi{V}_s^{(l\sigma)}=\bm{\chi}^{(l\sigma,q)}\odot\bi{C}^{(q)}.
    \label{eq:chi-tensor}
\end{equation}
For the modes $l\sigma\in\{1s,2s,2a\}$ the tensor $\bm{\chi}^{(l\sigma,q)}$ was given in our previous work \citep{turkFluctuatingHydrodynamicsAutophoretic2024}. Note, however, that the expansion coefficients of the activity, the concentration and the phoretic mobility have been altered slightly here, yielding
\begin{align}
    \bi{V}_s^{(1s)}&=\sum_{q=1}^{\infty}\tfrac{1}{w_q}\left[\tfrac{q+1}{4q^{2}-1}\bi{M}^{(q-1)}-\tfrac{q(q+1)}{2q+3}\bi{M}^{(q+1)}\right]\odot\bi{C}^{(q)},\nonumber\\[0pt]
     \bi{V}_s^{(2a)}&=\sum_{q=1}^{\infty}\tfrac{3}{4w_q}\tfrac{q}{2q+1}\,\bi{M}^{(q)}\times'\bi{C}^{(q)}, \nonumber\\[0pt]
     \bi{V}_s^{(2s)}&=\sum_{q=1}^{\infty}\tfrac{3}{w_q}\left[\tfrac{q+1}{(4q^{2}-1)(2q-3)}\bi{M}^{(q-2)}+\tfrac{3q}{(2q+1)(2q+3)}\bi{M}_{{\rm sym}}^{(q)}-\tfrac{q(q+1)(q+2)}{2q+5}\bi{M}^{(q+2)}\right]\odot\bi{C}^{(q)},
     \label{eq:slip-old}
\end{align}
where the cross product for irreducible tensors is defined as $(\boldsymbol{M}^{(q)}\times'\boldsymbol{C}^{(q)}){}_{\alpha}=\epsilon_{\alpha\beta\gamma}M_{\beta({\scriptscriptstyle Q-1})}^{(q)}C_{\gamma({\scriptscriptstyle Q-1})}^{(q)}$ and the symmetric and traceless product contracting $q-1$ indices is defined as $(\bi{M}_{{\rm sym}}^{(q)}\odot\bi{C}^{(q)})_{\alpha\beta}=\Delta_{\alpha\beta,\alpha'\beta'}^{(2)}M_{\alpha'({\scriptscriptstyle Q-1})}^{(q)}C_{\beta'({\scriptscriptstyle Q-1})}^{(q)}$. We have used the short-hand notation $Q=\gamma_1\gamma_2\dots\gamma_q$ for Cartesian indices.
It is apparent that for a uniform phoretic mobility distribution at the surface of the particle, i.e., $\beta=1$, we have $M_q=0\,\forall\,q\geq1$, so that $\bi{V}_s^{(2a)}$, corresponding to chemically induced self-rotation of the particle, vanishes. 

In order to take into account all slip modes generating long-ranged flows, we also include the quadrupolar mode
\begin{equation}
    \bi{V}_s^{(3s)}=\sum_{q=1}^{\infty}\tfrac{15}{w_q}\left[\tfrac{q+1}{(4q^{2}-1)(2q-3)(2q-5)}\bi{M}^{(q-3)}-\tfrac{q(q+1)(q+2)(q+3)}{2q+7}\bi{M}^{(q+3)}\right]\odot\bi{C}^{(q)}
    \label{eq:slip-new}
\end{equation}
in our analysis. We have not included the mode $(3a)$ corresponding to a rotlet dipole. For a system with cylindrical symmetry such as an axisymmetric particle near a plane interface, this mode vanishes. It is also worth noting that, since the phoretic slip is tangential to the surface of the particle, i.e., $\bm{n}\cdot\bm{v}_s=0$, the expansion coefficients $\bi{V}_s^{(l\sigma)}$ are not all independent. Here, a relevant relation arising from this is $\bi{V}_s^{(3t)}=-5\bi{V}_s^{(1s)}$. 

To summarise, given the particle's activity \eqref{eq:activity-coeffs}, its chemical interactions with its surroundings \eqref{eq:elastances} and the phoretic mobility distribution on its surface \eqref{eq:mobility-coeffs}, we can compute its phoretic slip \eqref{eq:slip-expansion} and thus, its dynamics via the equations of motion \eqref{eq:eom}.

\begin{figure}
  \centerline{\includegraphics[width=.6\columnwidth]{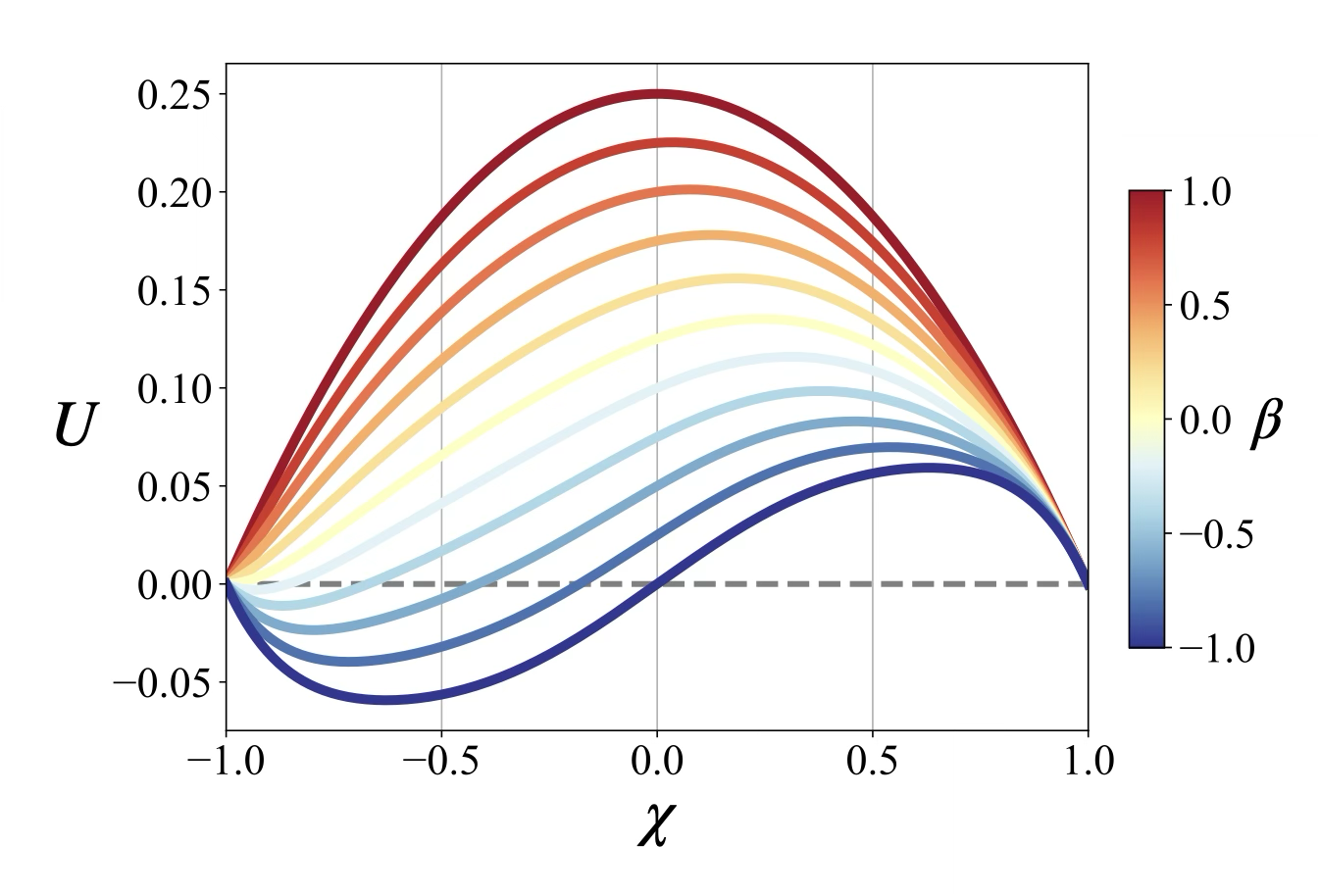}}
  \vspace{-3mm}
  \caption{
  Speed $U$ of an autophoretic particle in an unbounded fluid as a function of the size of its catalytic cap $\chi$ and the ratio of phoretic mobilities $\beta=\mu_i/\mu_c$, assuming $\mu_c>0$. 
  }
\label{fig:speed-unbounded}
\end{figure}

For example, in an unbounded fluid, the given activity and mobility coefficients yield the following particle speed:
\begin{equation}
    U=\tfrac{1}{64}(1-\chi^2)(8+5\chi-2\chi^3+5\chi^5 + \beta(8-5\chi + 2\chi^3 - 5\chi^5)),
\end{equation}
shown in figure \ref{fig:speed-unbounded} as a function of the cap size $\chi$ and the phoretic mobility ratio $\beta$.
For uniform phoretic mobility ($\beta=1$), this is an exact result \citep{michelinPhoreticSelfpropulsionFinite2014}. Assuming $\mu_c>0$ and $\beta\geq0$ ensures that the particle is chemo-repulsive, i.e., the particle behaves as an inert-side forward swimmer.

\section{Geometric cap model}
\label{sec:geometric-cap-model}

\begin{figure}
  \centerline{\includegraphics[width=.3\columnwidth]{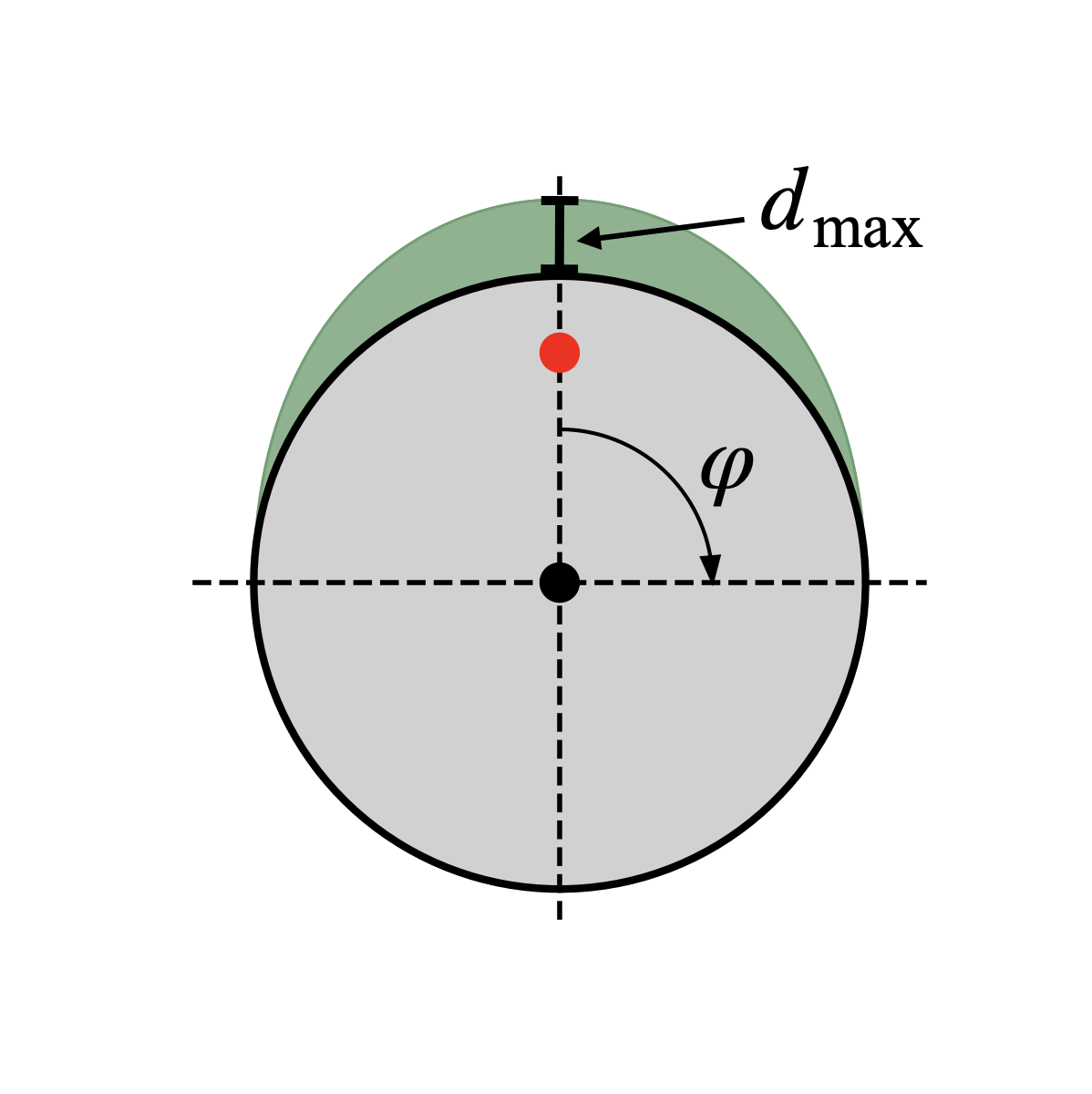}}
  \vspace{-5mm}
  \caption{
  Geometric cap model. The black dot indicates the centre of mass of the spherical particle (gray), while the red dot indicates the centre of mass of the catalytic cap (green). The contact angle $\varphi$ and the maximum thickness $d_{\rm max}$ of the catalytic cap are shown. The variation in thickness of the cap is given by equation \eqref{eq:thickness}.
  }
\label{fig:geometric-cap-model}
\end{figure}

To take into account cap-heaviness due to a mismatch between the particle's gravitational and geometric centres, we employ a geometric cap model defined by
\begin{equation}
    d(\alpha)=d_{\rm max}\cos^2\left(\frac{\pi}{2}\frac{\alpha}{\varphi}\right),
    \label{eq:thickness}
\end{equation}
where $d_{\rm max}$ is the maximum cap thickness at the pole relative to the particle radius, $\alpha\in[0,\varphi]$ and $\varphi$ is the contact angle of the catalytic cap as defined in figure \ref{fig:schematics}a. An example for a half-covered particle ($\varphi=\pi/2$) is shown in figure \ref{fig:geometric-cap-model}. The advantage of this model is that for a half-covered particle, the centre of mass (CoM) of the cap is at a distance $3/4$ from the centre of the unit sphere, matching an experimentally tested cap model by \citet{campbellGravitaxisSphericalJanus2013}. Furthermore, it is straightforwardly generalised to different sized catalytic caps. In general, the non-dimensionalised distance of the combined (particle and cap) CoM from the centre of the sphere is
\begin{equation}
    r_m=\begin{cases}
        \frac{3}{4}\frac{Kd}{Kd+2}\quad\text{for }\varphi=\pi/2,\\[5pt]
        \frac{3}{2}\frac{Kd(\pi^2-\varphi^2)(\pi^2-4\varphi^2-\pi^2\cos^2\varphi)}{(\pi^2-4\varphi^2)[4(\pi^2-\varphi^2)+3Kd(\pi^2-2\varphi^2-\pi^2\cos\varphi)]}\quad\text{else},
    \end{cases}
\end{equation}
where $K$ is the ratio of the buoyant volume densities of the cap and sphere. This model of the cap thickness is employed purely to determine the mass distribution. Chemically and hydrodynamically, the particle is assumed to be strictly spherical.

\section{Repulsive surface}
\label{sec:repulsive-surface}

\begin{figure}
  \centerline{\includegraphics[width=\columnwidth]{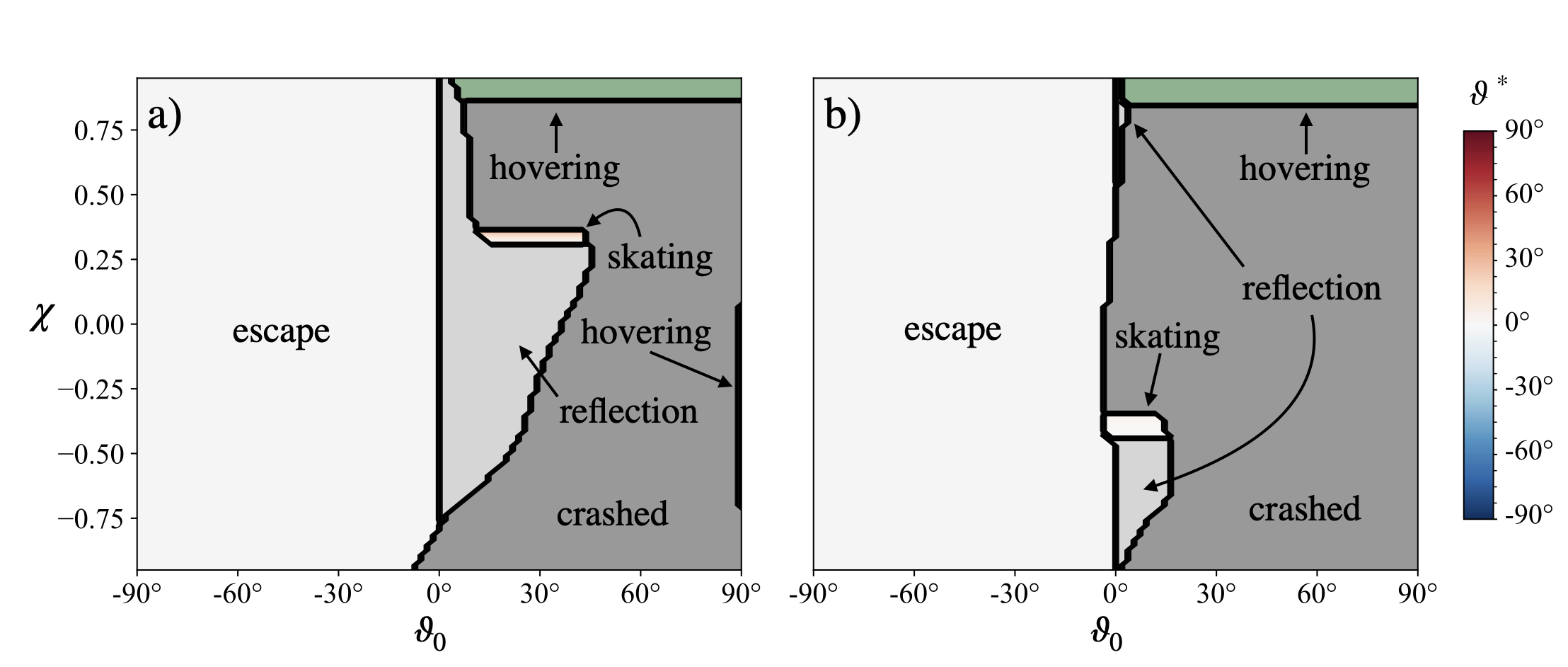}}
  \vspace{-2mm}
  \caption{
        Long-time behaviours of a neutrally buoyant Janus particle with a uniform phoretic mobility ($\beta=1$) as a function of its catalytic coverage $\chi$ (where $|\chi|\leq0.95$) and initial orientation $\vartheta_0$ near a rigid wall ($\lambda^f\rightarrow\infty$) in panel (a) and a free surface ($\lambda^f=0$) in panel (b) without the addition of a short-ranged repulsive particle-wall interaction. Either surface is assumed to be impermeable to the solutes ($\kappa_c=0$). The particle's initial height is $H_0=2$. 
        The swimmer has crashed into the wall for $H<1$. 
        If $H>30$ at any time, the particle is deemed to have escaped the wall for initial orientations away from the surface ($\vartheta_0<0$) and been reflected by the wall for initial orientations towards the surface ($\vartheta_0>0$). 
        For the skating state the steady tilt angle $\vartheta^*$ is indicated by the colour bar.
  }
\label{fig:pd-angles-noLJ}
\end{figure}

\begin{figure}
  \centerline{\includegraphics[width=\columnwidth]{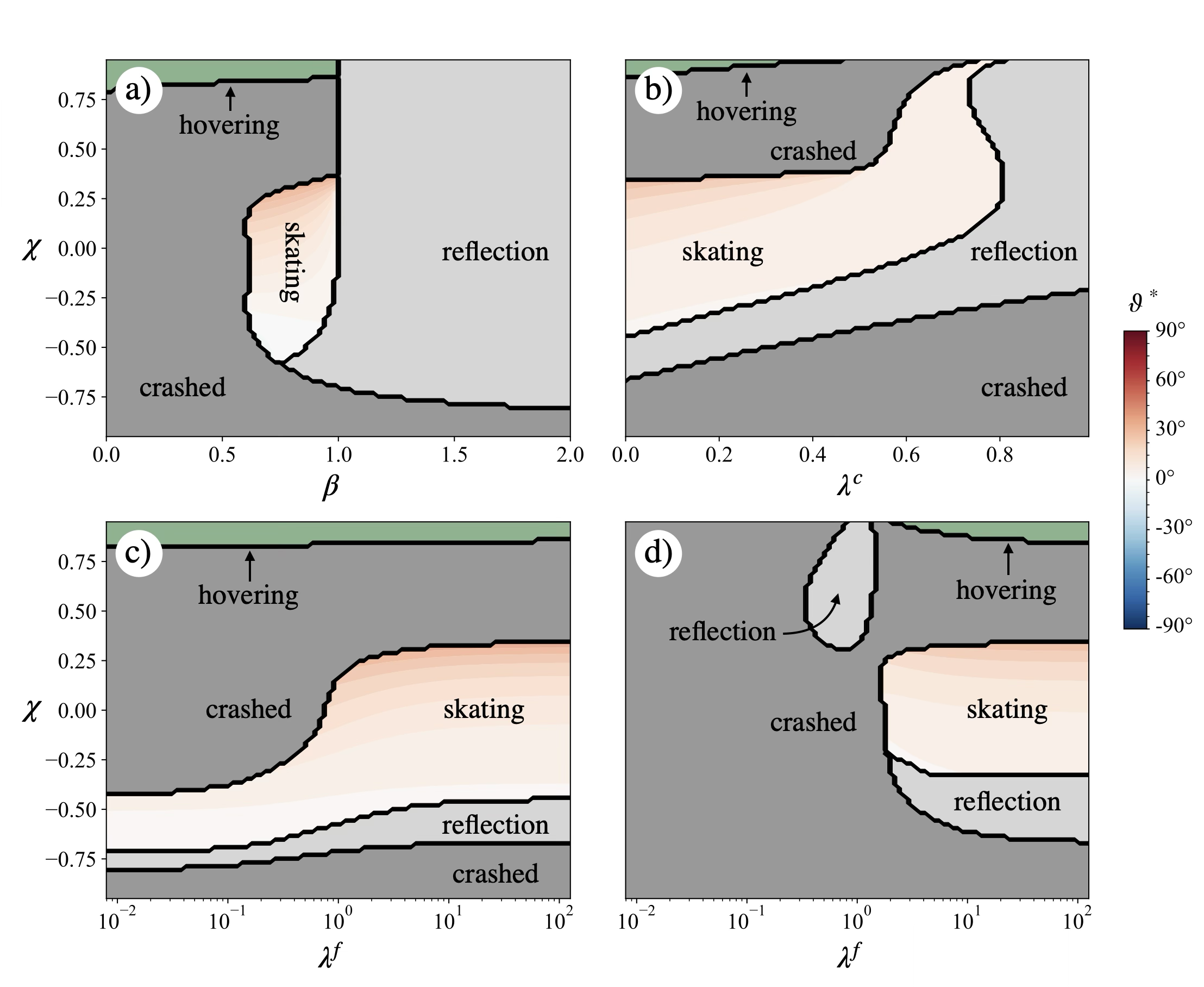}}
  \vspace{-5mm}
  \caption{
        Long-time behaviours of a neutrally buoyant Janus particle near various surfaces with initial conditions $H_0=2$ and $\vartheta_0=5^\circ$ without the addition of a short-ranged repulsive particle-surface interaction as a function of its cap size $\chi$ (where $|\chi|\leq0.95$) and other parameters.
        The swimmer has crashed into the boundary for $H<1$.
        The particle is deemed to have escaped the wall if $H>30$ at any time. 
        For the skating state, the steady tilt angle is indicated by the colour bar. 
        Panel (a) shows the phase diagram of the particle near an impermeable rigid wall ($\kappa_c=0$ and $\lambda^f\rightarrow\infty$) as a function of the phoretic mobility ratio $\beta$.
        For panels (b-d) we set $\beta=0.9$. 
        Panel (b) shows the phase diagram near a permeable rigid wall ($\kappa_c=1$, finite $\lambda^c$ and $\lambda^f\rightarrow\infty$) as a function of the diffusivity ratio $\lambda^c$. 
        Panel (c) shows the phase diagram near an impermeable fluid-fluid interface ($\kappa_c=0$ and finite $\lambda^f$) as a function of the viscosity ratio $\lambda^f$.
        Panel (d) shows the phase diagram near a permeable fluid-fluid interface ($\kappa_c=1$ and finite $\lambda^c=1/\lambda^f$) as a function of the viscosity ratio $\lambda^f$.
  }
\label{fig:pd-all-noLJ}
\end{figure}

\begin{figure}
  \centerline{\includegraphics[width=.6\columnwidth]{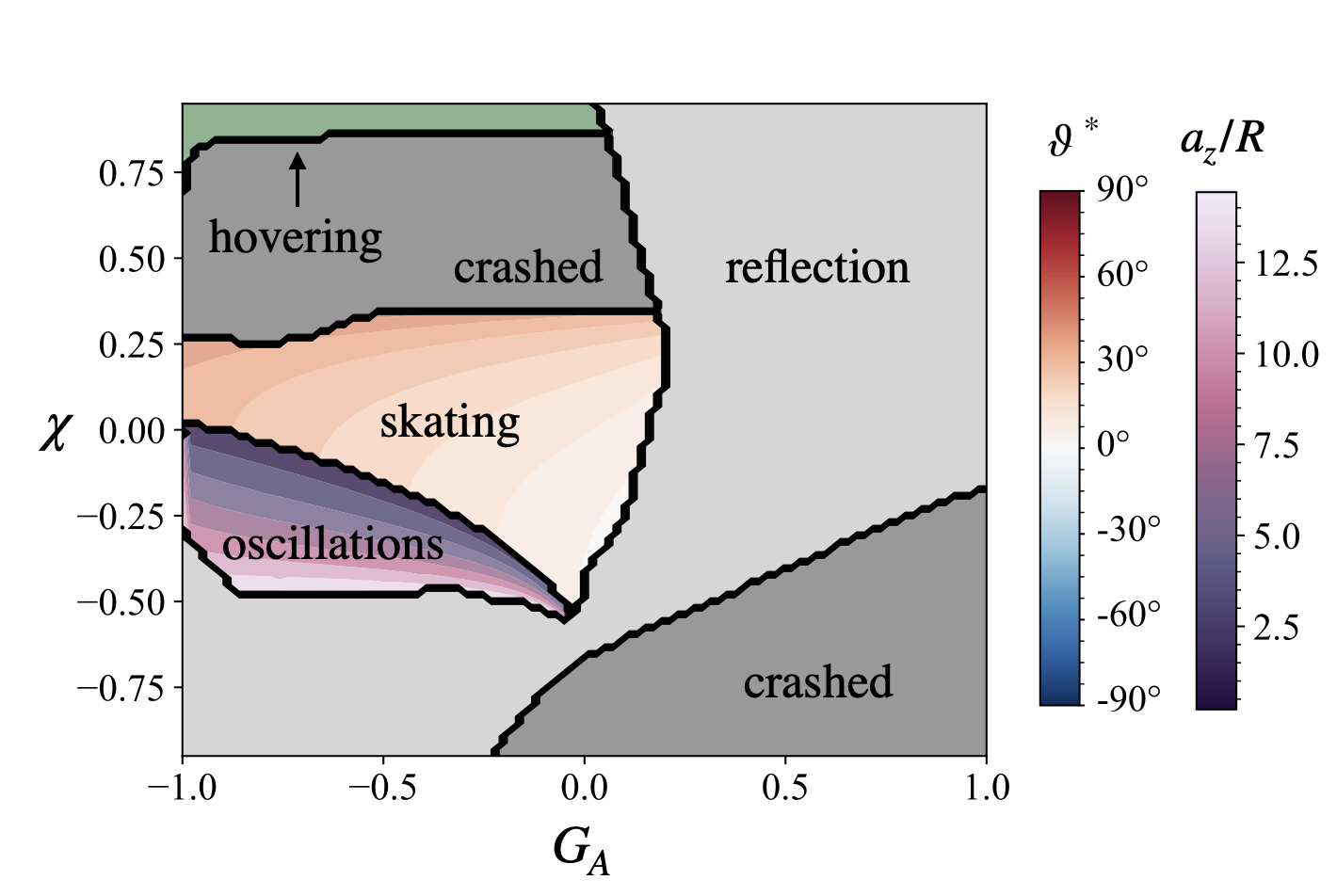}}
  \vspace{-2mm}
  \caption{
        Long-time behaviours of a buoyant Janus particle with $\beta=0.9$ near an impermeable rigid wall ($\kappa_c=0$ and $\lambda^f\rightarrow\infty$) under the influence of gravity and without the addition of a short-ranged repulsive particle-wall interaction.
        The particle's initial conditions are $H_0=2$ and $\vartheta_0=5^\circ$. 
        The phase diagram is shown as a function of the particle's cap size $\chi$ and its buoyancy $G_A$.
        The swimmer has crashed into the wall for $H<1$.
        The particle is deemed to have escaped the wall if $H>30$ at any time.
        For the skating and oscillating states, the skating angle $\vartheta^*$ and the relative amplitude of the oscillations in $z$-direction $a_z/R$ are indicated by the respective colour bars.
  }
\label{fig:pd-gravity-noLJ}
\end{figure}

The phase diagrams in the main text were created by integrating the equations of motion \eqref{eq:eom} over time, using a truncation of the generated chemo-hydrodynamic fields such that our equations are exact up to $\mathcal{O}\left (H^{-3}\right )$ for linear and $\mathcal{O}\left (H^{-4}\right )$ for angular interactions with the bounding surface. For small particle-boundary separations, therefore, our theory breaks down. To avoid the particle crashing into the boundary, we impose a shifted and truncated Lennard-Jones potential (WCA potential) between the particle and the boundary, given by
\begin{equation}
    W(H)=\begin{cases}
    \frac{\epsilon}{12}\left(\left(\frac{\sigma}{H}\right)^{12}-2\left(\frac{\sigma}{H}\right)^6 + 1\right),\quad\text{when}\quad H\leq\sigma,\\
    0,\quad\text{otherwise}
    \end{cases}
    \label{eq:wca}
\end{equation}
where $\epsilon$ is the strength of the potential and $\sigma/R$ is its relative reach. A stiff and short-ranged potential ($\epsilon=0.5$, $\sigma/R=1.1$) may be used to emulate a hard-core repulsion between the swimmer and the surface.
% We choose $\epsilon\approx0.35$ and $\sigma/R=1.1$, keeping the particle at a distance $H\geq 1.05$ from the boundary. 
% The factor $0.6\pi$ in the potential strength arises from the fact that the generated repulsive force is rescaled by $U/6\pi\eta R$, with a dimensional potential strength of $0.1$J ($=$ Joule).

Without the addition of this repulsive potential, we first illustrate the long-time behaviours of a neutrally buoyant Janus particle with uniform phoretic mobility ($\beta=1$) near a rigid wall ($\lambda^f\rightarrow\infty$) and near a free surface ($\lambda^f=0$), both impermeable to the solutes ($\kappa_c=0$), for the entire range of initial orientations and as a function of the particle coverage in figure \ref{fig:pd-angles-noLJ}. This matches previous results by \citet{ibrahimHowWallsAffect2016} for a rigid wall, where even without the addition of a repulsive potential, a skating state emerges around $\chi\approx0.32$ and a hovering state can be observed for coverages larger than $\chi\approx0.88$. The corresponding values for a stress-free surface are $\chi\approx-0.42$ and $\chi\approx0.85$. For the corresponding phase diagrams \emph{with} the addition of a repulsive potential, see figure \ref{fig:pd-angles}.

For a shallow initial particle orientation $\vartheta_0=5^\circ$ and without the addition of a short-ranged repulsive particle-wall interaction, we show the phase diagrams for the various surfaces considered in this paper in figure \ref{fig:pd-all-noLJ}. For the corresponding phase diagrams \emph{with} a repulsive potential, compare this with figure \ref{fig:pd-all-shallow}. It is clear that without a repulsive potential, for a large range of particle- and surface-properties, our theory breaks down/the particle crashes into the surface for the given initial conditions. However, chemo-hydrodynamic reflection, as well as robust skating and hovering states can still be observed in some cases.

The phase diagram for a cap-heavy Janus particle, affected by gravity, near a chemically impermeable rigid wall without the addition of a short-ranged repulsive particle-wall interaction is shown in figure \ref{fig:pd-gravity-noLJ} for an initial particle orientation $\vartheta_0=5^\circ$. For the corresponding phase diagrams \emph{with} a repulsive potential, compare this with figure  \ref{fig:pd-gravity-initial-angle}a.

\section{Varying the particle's initial conditions}
\label{sec:more-pds}

\begin{figure}
  \centerline{\includegraphics[width=\columnwidth]{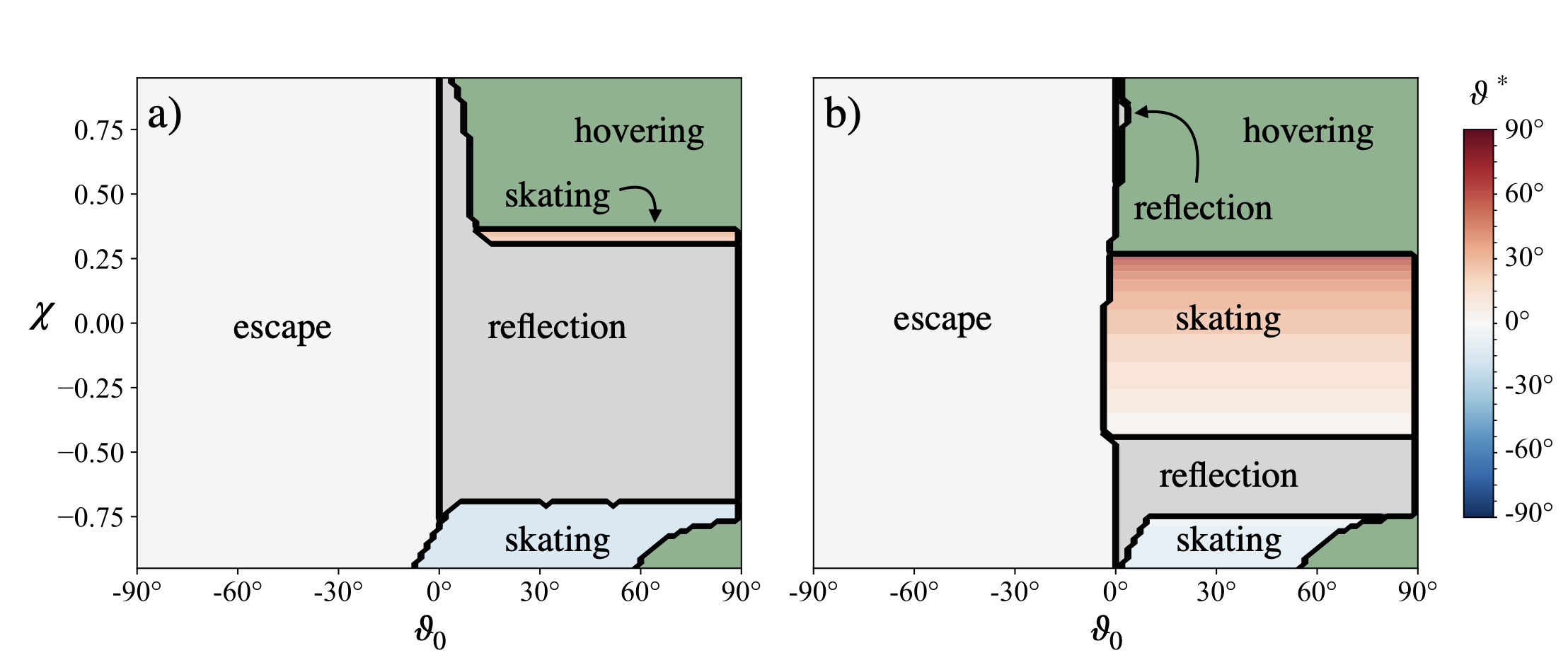}}
  \vspace{-2mm}
  \caption{
        Long-time behaviours of a neutrally buoyant Janus particle with a uniform phoretic mobility ($\beta=1$) as a function of its catalytic coverage $\chi$ (where $|\chi|\leq0.95$) and initial orientation $\vartheta_0$ near a rigid wall ($\lambda^f\rightarrow\infty$) in panel (a) and a free surface ($\lambda^f=0$) in panel (b). Either surface is assumed to be impermeable to the solutes ($\kappa_c=0$). The particle's initial height is $H_0=2$. 
        If $H>30$ at any time, the particle is deemed to have escaped the wall for initial orientations away from the surface ($\vartheta_0<0$) and been reflected by the wall for initial orientations towards the surface ($\vartheta_0>0$). 
        For the skating state the steady tilt angle $\vartheta^*$ is indicated by the colour bar. 
  }
\label{fig:pd-angles}
\end{figure}

\begin{figure}
  \centerline{\includegraphics[width=\columnwidth]{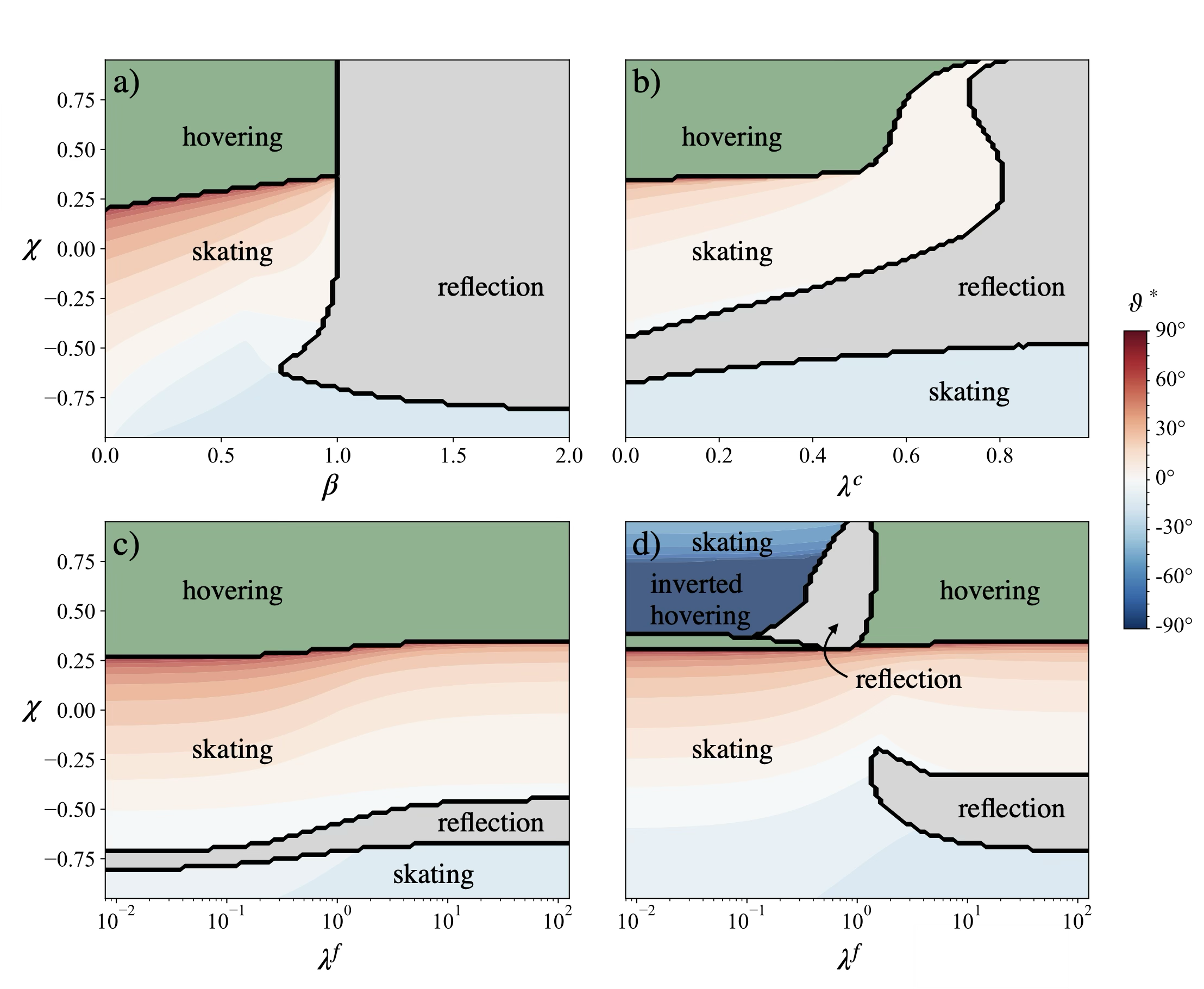}}
  \vspace{-5mm}
  \caption{
        Shallow initial orientation: Long-time behaviours of a neutrally buoyant Janus particle near various surfaces with initial conditions $H_0=2$ and $\vartheta_0=5^\circ$ as a function of its cap size $\chi$ (where $|\chi|\leq0.95$) and other parameters.
        The particle is deemed to have escaped the wall if $H>30$ at any time. 
        For the skating state, the steady tilt angle is indicated by the colour bar. 
        Panel (a) shows the phase diagram of the particle near an impermeable rigid wall ($\kappa_c=0$ and $\lambda^f\rightarrow\infty$) as a function of the phoretic mobility ratio $\beta$.
        For panels (b-d) we set $\beta=0.9$. 
        Panel (b) shows the phase diagram near a permeable rigid wall ($\kappa_c=1$, finite $\lambda^c$ and $\lambda^f\rightarrow\infty$) as a function of the diffusivity ratio $\lambda^c$. 
        Panel (c) shows the phase diagram near an impermeable fluid-fluid interface ($\kappa_c=0$ and finite $\lambda^f$) as a function of the viscosity ratio $\lambda^f$.
        Panel (d) shows the phase diagram near a permeable fluid-fluid interface ($\kappa_c=1$ and finite $\lambda^c=1/\lambda^f$) as a function of the viscosity ratio $\lambda^f$. The region of `inverted hovering' indicates a stationary fluid-pumping state in which the catalytic cap is turned towards the interface, i.e., $\vartheta^*=-90^\circ$.
  }
\label{fig:pd-all-shallow}
\end{figure}

\begin{figure}
  \centerline{\includegraphics[width=\columnwidth]{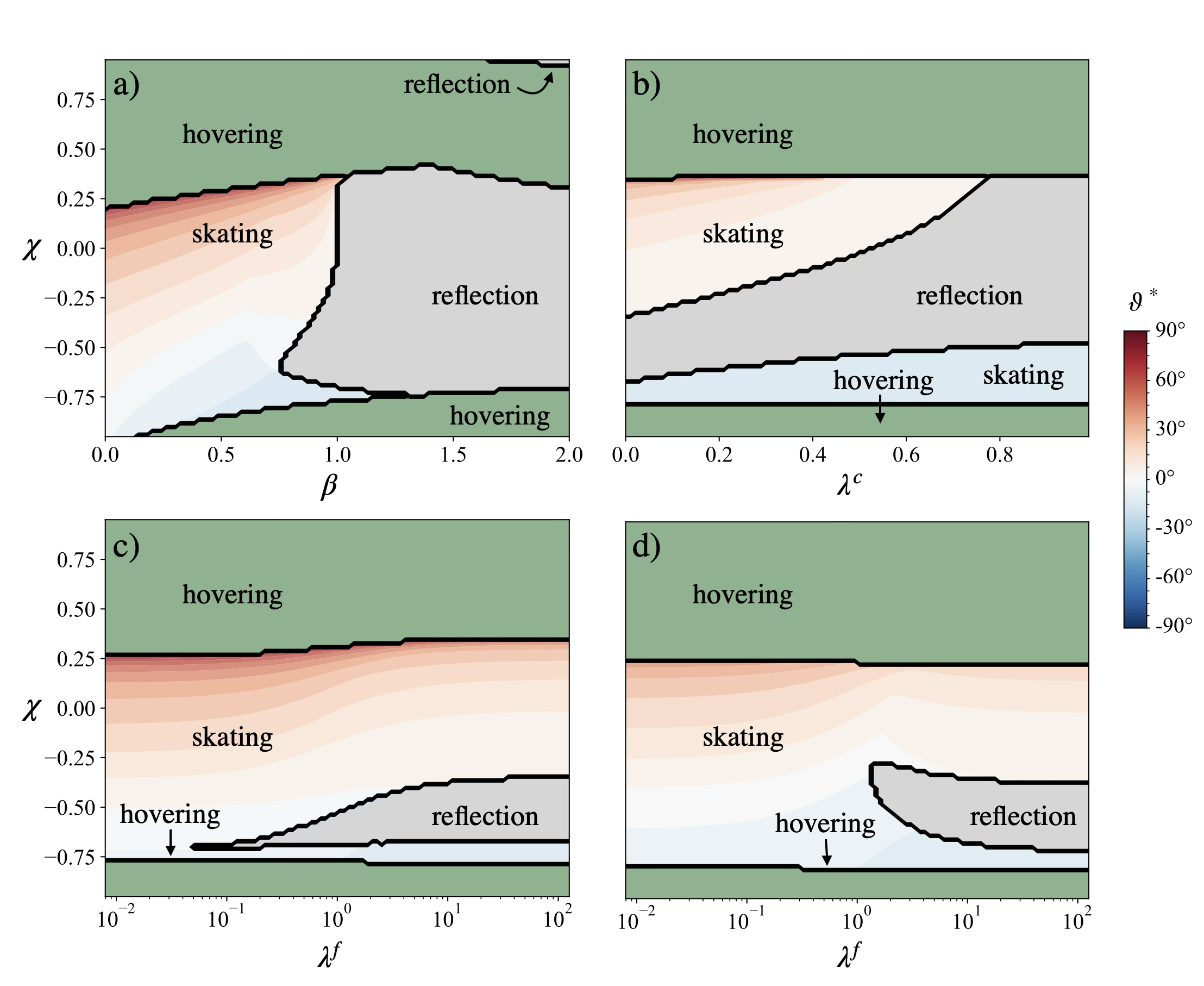}}
  \vspace{-5mm}
  \caption{
        Steep initial orientation: Long-time behaviours of a neutrally buoyant Janus particle near various surfaces with initial conditions $H_0=2$ and $\vartheta_0=85^\circ$ as a function of its cap size $\chi$ (where $|\chi|\leq0.95$) and other parameters.
        The particle is deemed to have escaped the wall if $H>30$ at any time. 
        For the skating state, the steady tilt angle is indicated by the colour bar. 
        Panel (a) shows the phase diagram of the particle near an impermeable rigid wall ($\kappa_c=0$ and $\lambda^f\rightarrow\infty$) as a function of the phoretic mobility ratio $\beta$.
        For panels (b-d) we set $\beta=0.9$. 
        Panel (b) shows the phase diagram near a permeable rigid wall ($\kappa_c=1$, finite $\lambda^c$ and $\lambda^f\rightarrow\infty$) as a function of the diffusivity ratio $\lambda^c$. 
        Panel (c) shows the phase diagram near an impermeable fluid-fluid interface ($\kappa_c=0$ and finite $\lambda^f$) as a function of the viscosity ratio $\lambda^f$.
        Panel (d) shows the phase diagram near a permeable fluid-fluid interface ($\kappa_c=1$ and finite $\lambda^c=1/\lambda^f$) as a function of the viscosity ratio $\lambda^f$.
  }
\label{fig:pd-all-steep}
\end{figure}

\begin{figure}
  \centerline{\includegraphics[width=\columnwidth]{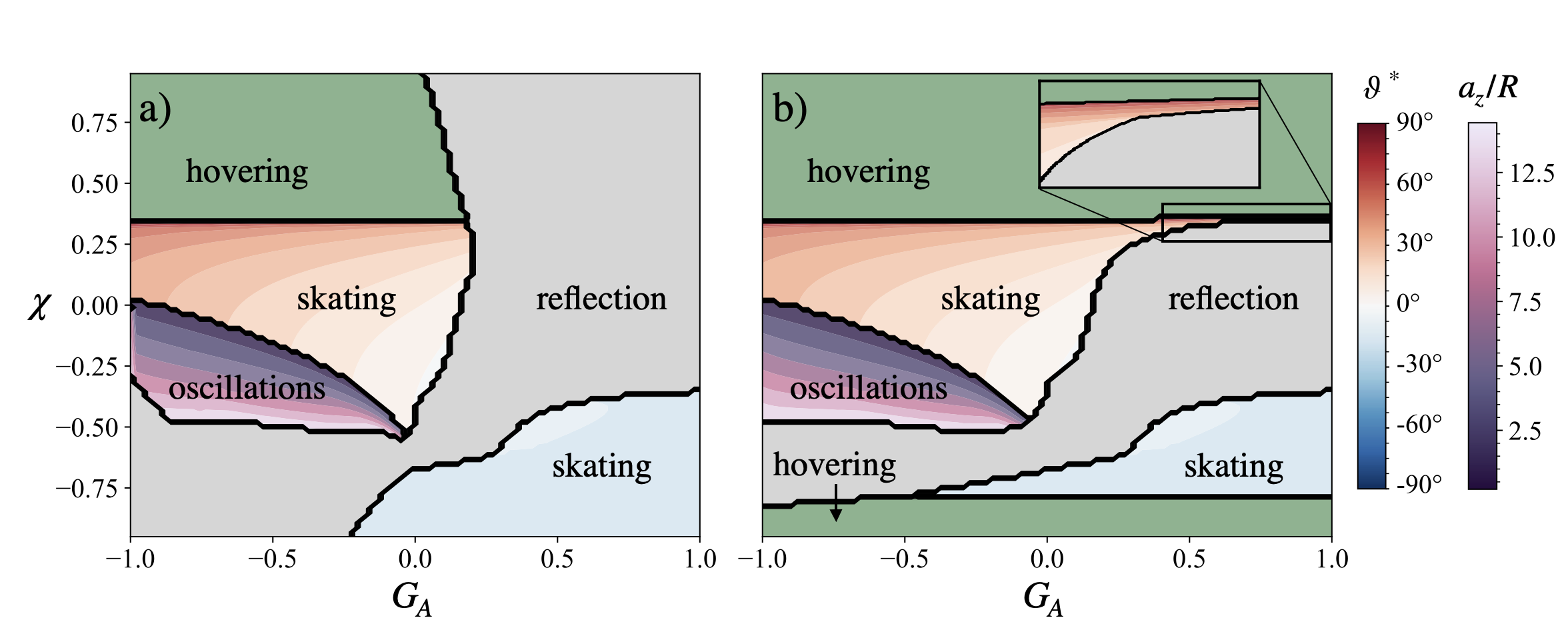}}
  \vspace{-2mm}
  \caption{
        Long-time behaviours of a buoyant Janus particle with $\beta=0.9$ near an impermeable rigid wall ($\kappa_c=0$ and $\lambda^f\rightarrow\infty$) under the influence of gravity.
        The particle's initial height above the interface is $H_0=2$. 
        The initial orientations are $\vartheta_0=5^\circ$ in panel (a) and $\vartheta_0=85^\circ$ in panel (b). The phase diagrams are shown as functions of the particle's cap size $\chi$ (where $|\chi|\leq0.95$) and its buoyancy $G_A$.
        The particle is deemed to have escaped the wall if $H>30$ at any time.
        For the skating and oscillating states, the skating angle $\vartheta^*$ and the relative amplitude of the oscillations in $z$-direction $a_z/R$ are indicated by the respective colour bars.
        The inset in panel (b) shows the detailed dynamics for $0.33<G_A<1$ and $0.26<\chi<0.38$.
  }
\label{fig:pd-gravity-initial-angle}
\end{figure}

In the main text and the phase diagrams in figures \ref{fig:pd-all} and \ref{fig:pd-gravity} we set the particle's initial conditions to be $H_0=2$ and $\vartheta_0=45^\circ$. Here, we discuss in more detail the effect of the particle's initial orientation, while keeping the initial height above the wall fixed. 

First, we illustrate the long-time behaviours of a neutrally buoyant Janus particle with uniform phoretic mobility ($\beta=1$) near a rigid wall ($\lambda^f\rightarrow\infty$) and near a free surface ($\lambda^f=0$), both impermeable to the solutes ($\kappa_c=0$), for the entire range of initial orientations and as a function of the particle coverage in figure \ref{fig:pd-angles}. 
This matches previous results by \cite{mozaffariSelfdiffusiophoreticColloidalPropulsion2016} and \cite{ibrahimHowWallsAffect2016} for a rigid wall qualitatively with the exception of two features. 
The skating and hovering regions for very small cap sizes have not been observed by other works. As discussed in the main text, we find that this region of the phase diagram is sensitive to the choice of potential between the particle and the wall, discussed in Appendix \ref{sec:repulsive-surface}.
\citet{ibrahimHowWallsAffect2016}, who used a similar truncation of chemo-hydrodynamic effects as is used here, do, however, also observe that the skating region for $\chi>0$ is narrow compared to results in the literature obtained via BEM or bispherical coordinates, the cause of which is likely to be two-fold. On the one hand, inevitably there will be approximation errors introduced by the truncation of the expansion of the surface fields. On the other hand, it is a known effect of external forces, such as the imposed repulsive particle-wall interaction, to force transitions from skating to stationary states \citep{mozaffariSelfdiffusiophoreticColloidalPropulsion2016}.

Next, we demonstrate that as long as $\vartheta_0>0$, varying the particle's initial orientation merely leads to a shift in some of the phase boundaries without the emergence of fundamentally novel long-time behaviours. For a particle initially oriented at a shallow angle ($\vartheta_0=5^\circ$), the phase diagrams corresponding to a variety of surfaces are shown in figure \ref{fig:pd-all-shallow}. Compared with the phase diagrams discussed in the main text ($\vartheta_0=45^\circ$), most phase boundaries are shifted only minimally. 
This is with the exception of a chemically permeable fluid-fluid interface (panel (d)) of small viscosity ratio $\lambda^f=1/\lambda^c\ll1$, where the long-time behaviour of the particle changes qualitatively compared to a steeper initial angle. 
As discussed in the main text, for $\lambda^f<1$ ($\lambda^c>1$), the permeable interface becomes chemically attractive to the catalytic cap, even resulting in a novel `inverted hovering' state. 
As in the hovering state, in this state the particle effectively acts as a stationary micro-pump for the fluid. 
However, in this case the catalytic cap faces the interface ($\vartheta=-90^\circ$) instead of facing away from it.
This distinctive behaviour is absent near impermeable surfaces.

For a steep initial orientation of the particle ($\vartheta_0=85^\circ$) the corresponding phase diagrams are shown in figure \ref{fig:pd-all-steep}. The only novel feature in these phase diagrams across all types of surfaces considered here is an emerging hovering state at very small cap sizes.

The corresponding results for a buoyant Janus particle near a chemically impermeable rigid wall, affected by gravity, are shown in figure \ref{fig:pd-gravity-initial-angle} for the two initial orientations $\vartheta_0=5^\circ$ and  $\vartheta_0=85^\circ$. 
Compared with the results discussed in the main text ($\vartheta_0=45^\circ$) the phase diagrams are only marginally altered.
For a shallow initial angle, particles tend to get reflected by the surface due to gravity rotating their caps towards the wall when $G_A>0$.
As in the case without gravity, for a steep initial angle, a hovering region emerges for small cap sizes.

\section{Chemo-hydrodynamic orientational balance}
\label{sec:angular-velocity}

\begin{figure}
  \centerline{\includegraphics[width=\columnwidth]{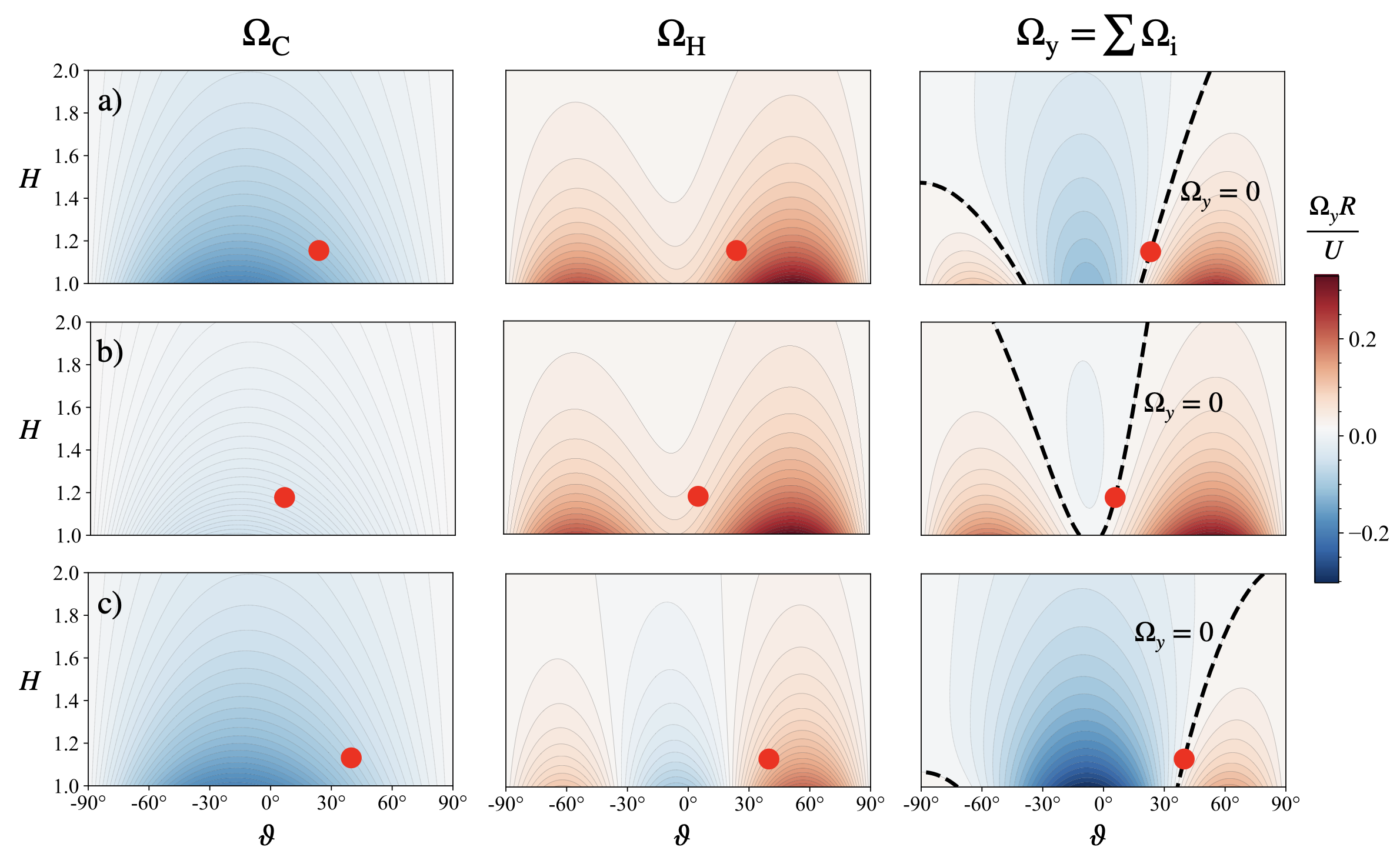}}
  \vspace{-2mm}
  \caption{
  Chemical ($\Omega_{\rm C}$) and hydrodynamic ($\Omega_{\rm H}$) contributions to a particle's ($\chi=0$, $\beta=0.5$) total angular velocity ($\Omega_y=\sum_i\Omega_i$) as described by equation \eqref{eq:angular-vel} for the cases of an impermeable rigid wall ($\kappa_c=0$ and $\lambda^f\rightarrow\infty$) in row (a), a permeable rigid wall ($\kappa_c=1$, $\lambda^c=0.5$ and $\lambda^f\rightarrow\infty$) in row (b) and an impermeable free surface ($\kappa_c=0$ and $\lambda^f=0$) in row (c). A dashed black line indicates a region of zero angular velocity on which the steady skating state (red dot) lies. According to the overlaid pseudo-colour map for the dimensionless angular velocity, red and blue indicate clockwise and anti-clockwise angular velocities, respectively. 
  }
\label{fig:angular-velocity}
\end{figure}

For the skating state, the intricate balance in the particle's fixed tilt angle can be understood by writing its angular velocity (here, in $y$-direction, see figure \ref{fig:schematics}) as the sum \citep{simmchenTopographicalPathwaysGuide2016}
\begin{equation}
    \Omega_y=\sum_i\Omega_i=\Omega_{\rm fs} + \Omega_{\rm C} + \Omega_{\rm H} + \Omega_{\rm CH} + \Omega_{\rm G}.
    \label{eq:angular-vel}
\end{equation}
In free space, i.e., without confining boundaries, the angular velocity is zero due to the particle's axisymmetry, $\Omega_{\rm fs}=0$. The terms $\Omega_{\rm C}$ and $\Omega_{\rm H}$ take into account chemical and hydrodynamic wall-interactions, respectively. The latter is comparable to the hydrodynamic wall-interactions of a squirmer. 
The term $\Omega_{\rm CH}$ contains higher order chemo-hydrodynamic couplings not included in the former two terms. This term decays faster than $H^{-3}$ in the relative distance from the surface and will therefore be ignored here. 
Finally, the term $\Omega_{\rm G}$ arises due to the particle's cap-heaviness when subjected to gravity. However, in the following we shall assume a neutrally buoyant particle, focusing instead on the chemical and hydrodynamic contributions to the particle's angular velocity only. 

In figure \ref{fig:angular-velocity} we compare this intricate balance of angular velocities for the cases of an impermeable rigid wall ($\lambda^f\rightarrow\infty$), a permeable rigid wall ($\lambda^f\rightarrow\infty$, $\lambda^c=0.5$) and an impermeable free surface ($\lambda^f=0$). It is clear that solute (chemical) interactions generally turn the catalytic cap away from the boundary, while hydrodynamic interactions tend to have the opposite effect, turning the cap towards the surface. These opposing effects lead to a line of zero angular velocity in the phase plane of the particle on which the steady skating state lies.  

Compared to an impermeable rigid wall, permeability and a finite diffusivity ratio $\lambda^c$ lead to weakened chemical wall-interactions, while the purely hydrodynamic interactions remain unchanged. In general, this leads to a shallower skating angle near permeable surfaces when compared to impermeable ones. 

In contrast, when considering a free surface that is impermeable to the solutes, the vanishing viscosity ratio $\lambda^f$ causes a reduction in the hydrodynamic torque generated by particle-surface interactions, while chemical effects remain unchanged. This leads to a steeper skating angle near a free surface when compared to an interface of finite viscosity ratio or a rigid wall.

\bibliographystyle{jfm}
%\bibliography{references}

\end{document}